\documentclass{elsart}
\usepackage{graphicx,amssymb,lineno,epsfig,amscd,subfigure}
\makeatletter
\def\elsartstyle{%
    \def\normalsize{\@setfontsize\normalsize\@xiipt{14.5}}
    \def\small{\@setfontsize\small\@xipt{13.6}}
    \let\footnotesize=\small
    \def\large{\@setfontsize\large\@xivpt{18}}
    \def\Large{\@setfontsize\Large\@xviipt{22}}
    \skip\@mpfootins = 18\p@ \@plus 2\p@
    \normalsize
}
\@ifundefined{square}{}{}
\makeatother

\newcommand{\bra}[1]{\langle #1|}
\newcommand{\ket}[1]{|#1\rangle}
\newcommand{\rf}[1]{(\ref{#1})}

\newcommand{\bea}{\begin{eqnarray}}
\newcommand{\eea}{\end{eqnarray}}
\newcommand{\be}{\begin{equation}}
\newcommand{\ee}{\end{equation}}

\begin{document}

\begin{frontmatter}
\title{Spectrum of the non-abelian phase in Kitaev's honeycomb lattice model}

\author{Ville Lahtinen}$^1$ \ead{ville.lahtinen@quantuminfo.org}, 
\author{Graham Kells}$^2$, 
\author{Angelo Carollo}$^3$, 
\author{Tim Stitt}$^4$,  
\author{Jiri Vala}$^2$ and
\author{Jiannis K. Pachos}$^1$ \ead{j.k.pachos@leeds.ac.uk} \ead[url]{http://quantum.leeds.ac.uk/$\sim$phyjkp/}

\address{$^1$Quantum Information Group, School of Physics and Astronomy, University of Leeds, Leeds LS2 9JT, UK}
\address{$^2$Department of Mathematical Physics, National University of Ireland, Maynooth, Ireland}
\address{$^3$Institute for Quantum Optics and Quantum Information of the Austrian Academy of Sciences, A-6020, Innsbruck, Austria}
\address{$^4$Irish Centre for High-End Computing, National University of Ireland, Galway, Ireland }

\begin{abstract}
The spectral properties of Kitaev's honeycomb lattice model are investigated
both analytically and numerically with the focus on the non-abelian phase
of the model. After summarizing the fermionization technique which maps
spins into free Majorana fermions, we evaluate the spectrum of
sparse vortex configurations and derive the interaction between two vortices as
a function of their separation. We consider the effect vortices can have on
the fermionic spectrum as well as on the phase transition between the
abelian and non-abelian phases. We explicitly demonstrate the
$2^n$-fold ground state degeneracy in the presence of $2n$ well separated vortices
and the lifting of the degeneracy due to their short-range
interactions. The calculations are performed on an infinite lattice.
In addition to the
analytic treatment, a numerical study of finite size systems is performed
which is in exact agreement with the theoretical considerations. The general spectral properties of the non-abelian phase are considered for
various finite toroidal systems.
\end{abstract}

\end{frontmatter}

\section{Introduction}
\label{intro}

Topological quantum computation~\cite{Freedman04,Kitaev,Pachos06-2,Preskill}
is certainly among the most exotic proposals for performing fault-tolerant
quantum information processing. This approach has attracted considerable
interest, since it is closely related to the problem of classifying
topologically ordered phases in various condensed matter systems. The
connection is provided by anyonic quasiparticles, which appear as states of
topologically ordered systems with non-trivial statistical properties. Some
of these anyon models can support universal quantum computation. Up to now,
no complete classification of topological phases exists in terms of their
physical properties or their computational power. This is due to
the small number of analytically treatable models that exhibit topological
behavior. The most studied arena is the celebrated fractional Quantum Hall
effect~\cite{Moore,Read} appearing in a two dimensional electron gas when it
is subject to a perpendicular magnetic field. 

Recently various two
dimensional lattice models exhibiting topological behavior have been
proposed~\cite{Doucot,Duan,Freedman05,Kitaev05,Levin,Yao,Fendley} that enjoy
analytic tractability. One such lattice proposal is the honeycomb model introduced by Alexei Kitaev
\cite{Kitaev05}. It consists of a two dimensional honeycomb lattice with
spins at its vertices subject to highly anisotropic spin-spin interactions.
This model has several remarkable features. It is exactly solvable and can
thus be studied analytically. For particular values of the couplings, the
model can be mapped to $Z_2$ gauge theory on a square lattice (the toric
code), which supports abelian anyons. This anyon model
has been employed for performing various quantum information
tasks~\cite{Kitaev}. When one adds an external magnetic
field, the model supports non-abelian Ising anyons.  Even though neither model supports
universal quantum computation, particular variations of the latter have been
considered for this purpose~\cite{Bravyi,Tewari}. One expects that when the
couplings of the honeycomb lattice model are varied, the system will undergo
a phase transition between the abelian and non-abelian phases. The existence
of the different phases is only argued in the original work~\cite{Kitaev05}
based on mathematical considerations and no rigorous presentation of the
transition is provided.

So far, the studies on Kitaev's honeycomb lattice model have concentrated on the abelian
phase~\cite{Chen,Pachos06,Schmidt}. Here
we present an extensive study of its spectral properties in the presence of
an external magnetic field. Solving the model for various sparse vortex
configurations gives us qualitative and quantitative results for the
behavior of the spectrum in the non-abelian phase. The study includes the
explicit demonstration of zero modes in the presence of well separated
vortices and the lifting of the degeneracy due to their short-range
interaction. These properties are subsequently connected to the properties
of the Ising anyon model giving direct evidence that the low energy
behavior of the non-abelian phase is indeed captured by this model. In
addition, we consider the stability of the different phases, which is of
importance when one is interested in physically realizing the
model~\cite{Micheli05}. The analytic calculations are supported by exact
numeric diagonalizations of finite size systems. The exact agreement between
the analytic and numeric solutions for these finite size systems is
demonstrated and the effect of an external effective magnetic field on finite size systems is discussed. Our work generalizes
the analysis in~\cite{Pachos06} performed by one of the authors, where only
the abelian phases in the limiting vortex-free and full-vortex cases were
considered.

The paper is organized as follows. In Section~\ref{Spectrum} we give an
overview of the honeycomb lattice model. There we outline the analytic
approach for solving the model for various vortex configurations by
employing Majorana fermionization. Section~\ref{Analytic} provides
explicitly the analytic solution for the limiting cases of vortex-free and
full-vortex configurations. These calculations are subsequently generalized
to sparse vortex configurations. Section~\ref{analysis} forms the main body
of our work. There we analyze in detail the behavior of the spectrum in
different parts of the phase space and study how it is modified due to the
presence of vortices. A connection to the Ising anyon model is provided.
In Section~\ref{Numerics} we study exact numerical diagonalization of
various finite size systems and show the equivalence with the analytic
results. Final remarks and conclusions are given in
Section~\ref{conclusions}.

\section{The spectrum of arbitrary vortex configurations}
\label{Spectrum}

\subsection{The honeycomb lattice model}

We briefly review here the honeycomb lattice model and its analytical
treatment as described in~\cite{Kitaev05}. The model is defined on a
honeycomb lattice $\Lambda$ with spins residing at each site. The sites are
bi-colored black and white such that $\Lambda = \Lambda_B \cup \Lambda_W$,
where $\Lambda_B$ and $\Lambda_W$ are two triangular sublattices.
\begin{figure}[t]
\begin{center}
\includegraphics*[width=12cm]{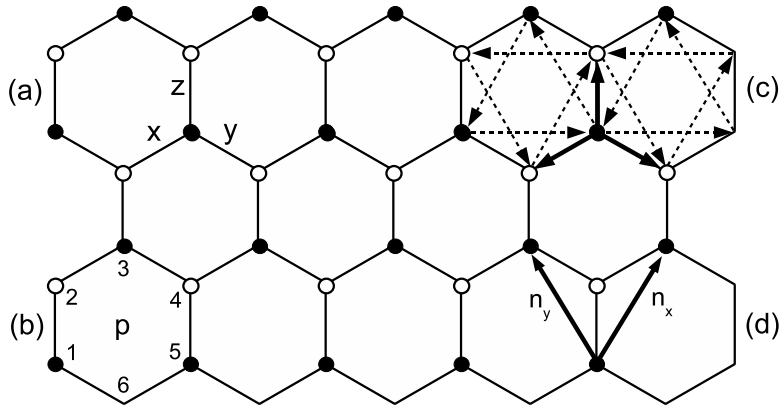}
\end{center}
\caption{\label{honeycomblattice}\small{The bi-colorable honeycomb lattice.
(a) Depending on their orientation the links are labelled as $x$, $y$ and
$z$. (b) A single plaquette $p$ with its sites enumerated. (c) Summation
convention for each elementary unit cell. Solid arrows indicate nearest
neighbor interactions along $x$- ,$y$- or $z$-links, whereas the dashed
arrows indicate next to nearest neighbor interactions originating from the
$K$-term \cite{Kitaev05}. (d) The elementary unit cell with lattice basis vectors
$\mathbf{n}_x$ and $\mathbf{n}_y$.}}
\end{figure}
We shall consider the following Hamiltonian
\begin{eqnarray} \label{H}
    H = -J_x \!\!\sum_{{\rm x-links}}\!\! \sigma^x_i \sigma^x_j -J_y
    \!\!\sum_{{\rm y-links}}\!\! \sigma^y_i \sigma^y_j -J_z
    \!\!\sum_{{\rm z-links}}\!\! \sigma^z_i \sigma^z_j - K
    \sum_{i,j,k} \sigma^x_i \sigma^y_j \sigma^z_k,
\end{eqnarray}
where $J_x, J_y$ and $J_z$ are positive coupling strengths along the $x$-,
$y$- and $z$-links, respectively, as shown in Figure
\ref{honeycomblattice}(a). The three-spin interaction, or the $K$-term, on
the right hand side can be obtained from a perturbative expansion when we apply a weak (Zeeman) magnetic field of the
form $H_h =
\mathbf{h}\cdot\mbox{\boldmath{$\sigma$}}$. In this case $K$ is given by $K
\approx \frac{h_x h_y h_z}{J^2}$, and this model is assumed to approximate
the one with a Zeeman term when $K \ll J_x,J_y,J_z$. Only this third order
term in the perturbative expansion will be of interest to us, since it is
the lowest order term breaking time reversal invariance~\cite{Kitaev05}. The
summations in the effective magnetic field term run over site triples such
that every plaquette $p$ contributes the six terms
\begin{eqnarray}
K(\sigma^z_1 \sigma^y_2 \sigma^x_3+\sigma^x_2 \sigma^z_3
\sigma^y_4+\sigma^y_3 \sigma^x_4 \sigma^z_5+\sigma^z_4 \sigma^y_5
\sigma^x_6+\sigma^x_5 \sigma^z_6 \sigma^y_1+\sigma^y_6 \sigma^x_1
\sigma^z_2).
\end{eqnarray}
The sites of a single plaquette have been enumerated as shown in
Figure~\ref{honeycomblattice}(b). The Hamiltonian~\rf{H}
commutes with the plaquette operators defined by
\begin{eqnarray} \label{w}
    \hat{w}_p = \sigma^x_1 \sigma^y_2 \sigma^z_3
    \sigma^x_4 \sigma^y_5 \sigma^z_6, \qquad \prod_p \hat{w}_p = I,
\end{eqnarray}
where the product is taken over all plaquettes of a compact lattice and $I$
is the identity operator. The eigenvalue $w_p =
- 1$ is interpreted as having a vortex on plaquette $p$. The constraint in \rf{w} implies that the vortices always come in
pairs. Since $w_p$ are conserved quantities, one can fix the underlying
vortex configuration and consider the Hamiltonian over this sector. This
would not be possible when the usual Zeeman term were employed, since it
does not commute with the plaquette operators. Thus, the magnetic field
induces hopping of the vortices between neighboring plaquettes and their
number is not necessarily conserved.

The Hamiltonian can be diagonalized by representing the spin operators with
Majorana fermions. Following~\cite{Kitaev05,Pachos06} we introduce two
fermionic modes $a_1$ and $a_2$ residing at each lattice site. The corresponding Majorana fermions are given by
\begin{eqnarray} \label{majoranas}
    c \equiv a_1 + a_1^\dagger, \quad b^x \equiv
    \frac{a_1 - a_1^\dagger}{i},
    \quad b^y \equiv a_2 + a_2^\dagger, \quad b^z \equiv
    \frac{a_2 - a_2^\dagger}{i}.
\end{eqnarray}
We encode the spin at each site at the subspace where both of the fermionic modes are either empty or full.
In terms of the four Majorana fermions, this means that we need to project down to
a two dimensional subspace, the physical space $\mathcal{L}$, by employing
the projector $D = b^x b^y b^z c$, i.e. $\ket{\Psi} \in \mathcal{L}
\Leftrightarrow D \ket{\Psi} = \ket{\Psi}$. The representation of the spin
matrices at site $i$ is then given in terms of the Majorana fermions by
$\sigma^\alpha_i = i b^\alpha_i c_i$, which satisfy the Pauli algebra when
restricted in $\mathcal{L}$ (note that $[D_i,\sigma_j^\alpha]=0$). It
follows that
\begin{eqnarray} \label{majorana_transforms}
    \sigma^\alpha_i \sigma^\alpha_j  =  -i \hat{u}_{ij} c_i c_j \quad \textrm{and} \quad
    \sigma^x_i \sigma^y_j \sigma^z_k  =
    -i \hat{u}_{ik} \hat{u}_{jk} c_i c_j,
\end{eqnarray}
where we have defined the operators
\begin{eqnarray} \label{u}
    \hat{u}_{ij} = ib^\alpha_i b^\alpha_j, \quad \left( \hat{u}_{ij} = -\hat{u}_{ji}, \quad \hat{u}_{ij}^2 = 1, \quad \hat{u}_{ij}^\dagger = \hat{u}_{ij} \right),
\end{eqnarray}
with $\alpha = x,y,z$ depending whether $i \in \Lambda_B$ and $j \in \Lambda_W$ are connected by a $x$-,
$y$- or $z$-link, respectively. Consequently, Hamiltonian \rf{H} takes the
form
\begin{eqnarray} \label{H_majorana}
     H = \frac{i}{4} \sum_{i,j \in \Lambda} \hat{A}_{ij} c_{i} c_{j},
     \qquad \hat{A}_{ij} = 2J_{ij} \hat{u}_{ij} + 2K \sum_{k}
     \hat{u}_{ik} \hat{u}_{jk}.
\end{eqnarray}
The explicit appearance of the constraint $D$ in \rf{majorana_transforms}
has been omitted, as we consider only operations in the physical subspace
$\mathcal{L}$. The couplings are given by $J_{ij}= J_x, J_y$ or $J_z$. The
summations in the Hamiltonian~\rf{H_majorana} are most conveniently
expressed pictorially (see Figure~\ref{honeycomblattice}(c)). The solid
lines correspond to nearest neighbor (the first term of $A_{ij}$ in
\rf{H_majorana}) and the dashed lines to next to nearest neighbor summation
(the second term of $A_{ij}$ in \rf{H_majorana}). The antisymmetry of the
$u$'s in \rf{u} is taken into account by using a convention such that one
assigns an overall $+$ ($-$) to every term involving sites $i \in \Lambda_B$ and
$j \in \Lambda_W$ when the arrow points from $i$ to $j$ ($j$ to $i$). If two
sites are not connected by an arrow the corresponding $A_{ij}$ element is
zero.

The plaquette
operators~\rf{w} can be written in terms of the $\hat{u}$'s as
\begin{eqnarray}
    \hat{w}_p = \prod_{i,j \in \partial p} \hat{u}_{ij}, \qquad i \in \Lambda_B, \quad j \in \Lambda_W,
\end{eqnarray}
where $\partial p$ denotes the boundary of plaquette $p$. Also, one can
check that $[H, \hat{u}_{ij}] = 0$. These observations imply that after
performing the fermionization, the underlying vortex configuration can be
fixed by specifying the eigenvalues $u_{ij} = \pm 1$ of the operators
$\hat{u}_{ij}$ on every link of the model. The eigenvalue $u_{ij} = -1$
means that there is a string passing through the link $ij$ that either
connects two vortices or belongs to a loop. The locations $p$ of these
vortices are determined by the eigenvalues $w_p = -1$ of the plaquette
operators.

\subsection{Solution for periodic vortex configurations}
\label{periodic}

\begin{figure}[t]
\begin{center}
\includegraphics*[width=8cm]{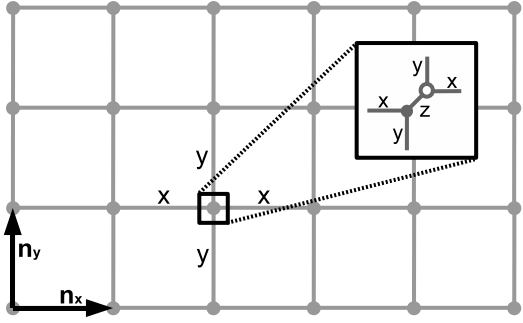}
\end{center}
\caption{\small{ The square lattice representation of the honeycomb lattice.
Every vertex contains one black and one white site connected by a single $z$-link of the initial lattice. The basis vectors $\mathbf{n}_x$ and $\mathbf{n}_y$ (Figure \ref{honeycomblattice}(d)) become orthogonal unit vectors along the $x$- and $y$-links.}}
\label{unitcell}
\end{figure}

Let us now consider in more detail the form of Hamiltonian~\rf{H_majorana}
for various periodic vortex configurations and its diagonalization by using
Fourier transform. Without affecting the physics of our system we shall
deform the original honeycomb lattice to a square lattice by taking the
length of $z$-links to zero and choosing the lattice basis vectors to be
$\mathbf{n}_x = (1,0)$ and $\mathbf{n}_y = (0, 1)$. The resulting square
lattice is shown in Figure~\ref{unitcell}.

First we determine the unit cell of our periodic vortex lattice. The
simplest possible choice contains a $z$-link of the honeycomb lattice, or in
other words a single site on the square lattice. We refer to this choice of
unit cell as the elementary cell.
In order to employ Fourier transform in diagonalizing the Hamiltonian the
underlying vortex configuration must be periodic with respect to the choice
of the unit cell. For the elementary cell there is only one such
configuration - the vortex-free configuration \cite{Kitaev05}. The
full-vortex configuration, i.e. a vortex on every plaquette, can be solved
by taking two elementary cells in which $\hat u$ alternates its sign along
one direction~\cite{Pachos06}. However, our aim is to go beyond these two
limiting cases and consider arbitrary sparse vortex configurations. To do
this, we define a generalized $(M, N)$-unit cell, which contains $MN$
elementary cells. On the square lattice the basis vectors take the simple
form
\begin{eqnarray} \label{sparsebasis}
    \mathbf{v}_x = M \mathbf{n}_x =  (M, 0), \qquad \mathbf{v}_y =
    N \mathbf{n}_y = (0, N).
\end{eqnarray}
An arbitrary site $i$ on the original honeycomb lattice can be labeled by
the triplet $i \to (\mathbf{r},k,\lambda)$, where $\mathbf{r}$ is a vector
indicating the location of the unit cell, the index pair ${\bf k} = (m,n),
1\leq m \leq M, 1\leq n \leq N,$ specifies a particular elementary cell
inside the generalized unit cell and
$\lambda=1,2$ is an index specifying whether the site belongs to $\Lambda_B$
or $\Lambda_W$. If $C$ denotes the number of unit cells, there are
altogether $2MNC$ sites on the lattice.

Using this notation the Hamiltonian \rf{H_majorana} can be written as
\begin{eqnarray} \label{H_majorana2}
     H = \frac{i}{4} \sum_{r,v} \sum_{k,l}^{(M,N)}
     \sum_{\lambda, \mu = 1}^2 A_{\lambda k \mu l}
     (\mathbf{v}) c_{\lambda k} (\mathbf{r}) c_{\mu l} (\mathbf{r+v}),
\end{eqnarray}
where the vector $\mathbf{v}$ is summed over all linear combinations of the
lattice basis vectors~\rf{sparsebasis}. Since $[H,\hat{u}]=0$, the operators
$\hat{u}$ \rf{u} appearing in $A_{\lambda k \mu l}$ \rf{H_majorana} have
been replaced with their eigenvalues $u = \pm 1$. The antisymmetry of the
operators $\hat{u}$ is included in the summation convention
(Figure~\ref{honeycomblattice}(c)). The properly normalized Fourier
transformation of the operators $c_{\lambda k}$ is given by
\begin{eqnarray}
    c_{\lambda k}(\mathbf{r}) & = & \sqrt{\frac{2}{C}}
    \int_{-\pi/M}^{\pi/M} \frac{d p_x}{\sqrt{2\pi/M}}
    \int_{-\pi/N}^{\pi/N} \frac{d p_y}{\sqrt{2\pi/N}}
    \ e^{i\mathbf{p}\cdot \mathbf{r}} c_{\lambda k}(\mathbf{p}).
\end{eqnarray}
Substituting this into \rf{H_majorana2} we obtain the canonical form
\begin{eqnarray} \label{H2}
    H & = & \frac{1}{2} \int_{-\pi/M}^{\pi/M} \frac{d p_x}{2\pi/M}
    \int_{-\pi/N}^{\pi/N} \frac{d p_y}{2\pi/N}
    \left( \begin{array}{c} \mathbf{c}_1(\mathbf{p})
    \\ \mathbf{c}_2(\mathbf{p}) \end{array} \right)^{\dagger} \left( \begin{array}{cc} A_{11}(\mathbf{p}) &
    A_{12}(\mathbf{p}) \\ A_{21}(\mathbf{p}) & A_{22}(\mathbf{p})
    \end{array}\right) \left( \begin{array}{c} \mathbf{c}_1(\mathbf{p})
    \\ \mathbf{c}_2(\mathbf{p}) \end{array} \right),
\end{eqnarray}
where $\mathbf{c}^{\dagger}_{\lambda}(\mathbf{p})=
(c_{\lambda(1,1)}^{\dagger}(\mathbf{p}),\ldots,c_{\lambda(M,N)}^{\dagger}(\mathbf{p}))$, and
$A_{\lambda \mu}(\mathbf{p})$ are matrices with elements $[A_{\lambda
\mu} (\mathbf{p})]_{kl} = \sum_v iA_{\lambda k \mu l}(\mathbf{v})
e^{-i\mathbf{p} \cdot \mathbf{v}}$. This Hamiltonian is a generalization of
the one obtained in~\cite{Kitaev05} with the exception that the single
entries of the $2 \times 2$ Hamiltonian are replaced here with $MN \times
MN$ matrices.

The off-diagonal blocks correspond to nearest-neighbor interactions. The
non-vanishing elements of the Hamiltonian for arbitrary $(M, N)$-unit cells
are given by
\begin{eqnarray} \label{A12}
\begin{array}{llll}
    \mathbf{c}^{\dagger}_1 A_{12} \mathbf{c}_2 = 2i( & + u_{k,k} & J_z & c_{1,k}^{\dagger}c_{2,k}^{\ } \\
    \ & + u_{k,k-n_x} & J_x e^{i\delta(m-1)\mathbf{p} \cdot \mathbf{v}_x} & c_{1,k}^{\dagger} c_{2,k-n_x}^{\ } \\
    \ & + u_{k,k-n_y} & J_y e^{i\delta(n-1)\mathbf{p} \cdot \mathbf{v}_y} & c_{1,k}^{\dagger} c_{2,k-n_y}^{\ }),
\end{array}
\end{eqnarray}
\begin{eqnarray} \label{A21}
\begin{array}{llll}
    \mathbf{c}^{\dagger}_2 A_{21} \mathbf{c}_1 = 2i( & - u_{k,k} & J_z & c_{2,k}^{\dagger}c_{1,k}^{\ } \\
    \ & - u_{k,k+n_x} & J_x e^{-i\delta(m-M)\mathbf{p} \cdot \mathbf{v}_x} & c_{2,k}^{\dagger} c_{1,k+n_x}^{\ } \\
    \ & - u_{k,k+n_y} & J_y e^{-i\delta(n-N)\mathbf{p} \cdot \mathbf{v}_y} & c_{2,k}^{\dagger} c_{1,k+n_y})^{\ },
\end{array}
\end{eqnarray}
where the addition in the indices ${\bf k}=(m,n)$ is understood $(m \bmod
M,n \bmod N)$ and $\delta (x) = 1$ for $x=0$ and $\delta (x) = 0$ otherwise. The diagonal blocks
correspond to next-to-nearest neighbor couplings and are given by
\begin{eqnarray} \label{A11}
\begin{array}{rlll}
    \mathbf{c}^{\dagger}_1 A_{11} \mathbf{c}_1 = 2iK( & + u_{k,k+n_y}^{k} & e^{-i\delta(n-N)\mathbf{p}\cdot \mathbf{v}_y} & c_{1,k}^{\dagger}c_{1,k+n_y}^{\ } \\
    \ & - u_{k,k-n_x+n_y}^{k-n_x} & e^{i\delta(m-1)\mathbf{p} \cdot \mathbf{v}_x} e^{-i\delta(n-N)\mathbf{p} \cdot \mathbf{v}_y} & c_{1,k}^{\dagger} c_{1,k-n_x+n_y}^{\ } \\
    \ & - u_{k,k+n_x}^{k} & e^{-i\delta(m-M)\mathbf{p} \cdot \mathbf{v}_x} & c_{1,k}^{\dagger} c_{1,k+n_x}^{\ } \\
    \ & + u_{k,k+n_x-n_y}^{k-n_y} & e^{-i\delta(m-M)\mathbf{p} \cdot \mathbf{v}_x} e^{i\delta(n-1)\mathbf{p} \cdot \mathbf{v}_y} & c_{1,k}^{\dagger}c_{1,k+n_x-n_y}^{\ } \\
    \ & + u_{k,k-n_x}^{k-n_x} & e^{i\delta(m-1)\mathbf{p} \cdot \mathbf{v}_x} & c_{1,k}^{\dagger} c_{1,k-n_x}^{\ } \\
    \ & - u_{k,k-n_y}^{k-n_y} & e^{i\delta(n-1)\mathbf{p} \cdot \mathbf{v}_y} & c_{1,k}^{\dagger} c_{1,k-n_y}^{\ }).
\end{array}
\end{eqnarray}
\begin{eqnarray} \label{A22}
\begin{array}{rlll}
    \mathbf{c}^{\dagger}_2 A_{22} \mathbf{c}_2 = 2iK( & -u_{k,k+n_y}^{k+n_y} & e^{-i\delta(n-N)\mathbf{p}\cdot \mathbf{v}_y} & c_{2,k}^{\dagger}c_{2,k+n_y}^{\ } \\
    \ & + u_{k,k-n_x+n_y}^{k+n_y} & e^{i\delta(m-1)\mathbf{p} \cdot \mathbf{v}_x} e^{-i\delta(n-N)\mathbf{p} \cdot \mathbf{v}_y} & c_{2,k}^{\dagger} c_{2,k-n_y+n_y}^{\ } \\
    \ & + u_{k,k+n_x}^{k+n_x} & e^{-i\delta(m-M)\mathbf{p} \cdot \mathbf{v}_x} & c_{2,k}^{\dagger} c_{2,k+n_x}^{\ }, \\
    \ & - u_{k,k+n_x-n_y}^{k+n_x} & e^{-i\delta(m-M)\mathbf{p} \cdot \mathbf{v}_x} e^{i\delta(n-1)\mathbf{p} \cdot \mathbf{v}_y} & c_{2,k}^{\dagger}c_{2,k+n_x-n_y}^{\ } \\
    \ & - u_{k,k-n_x}^{k} & e^{i\delta(m-1)\mathbf{p} \cdot \mathbf{v}_x} & c_{2,k}^{\dagger} c_{2,k-n_x}^{\ } \\
    \ & + u_{k,k-n_y}^{k} & e^{i\delta(n-1)\mathbf{p} \cdot \mathbf{v}_y} & c_{2,k}^{\dagger} c_{2,k-n_y}^{\ }),
\end{array}
\end{eqnarray}
where we have used the short-hand notation $u_{k,l}^{j} \equiv u_{k,j}
u_{j,l}$. Both $A_{11}$ and $A_{22}$ are Hermitian, which can be checked
by taking first Hermitian conjugates and subsequently shifting the indices
accordingly. Likewise, one can check the relations,
\begin{eqnarray}  \label{A12rel}
    A_{12} = A_{21}^{\dagger} \qquad \textrm{and} \qquad A_{22} = -A_{11}^T.
\end{eqnarray}
that guarantee the Hermiticity of $A$.

The expressions derived above give the most general expression for the
Hamiltonian of the honeycomb lattice model. To proceed with the
diagonalization, one needs to specify the underlying vortex configuration,
i.e. the values of $u$ on each link. Since all
bi-colorable Hamiltonians have a double spectrum \cite{Kitaev05}, we know
that the diagonalization of \rf{H2} will give the general form
\begin{eqnarray} \label{H_diag}
    H = \int_{-\pi/M}^{\pi/M} \frac{d p_x}{2\pi/M} \int_{-\pi/N}^{\pi/N} \frac{d p_y}{2\pi/N} \left( \sum_{i=1}^{MN} |\epsilon_i(\mathbf{p})| b_i^{\dagger} b_i - \sum_{i=1}^{MN} \frac{|\epsilon_i(\mathbf{p})|}{2} \right),
\end{eqnarray}
where $b_i$ are $MN$ fermionic operators and $\epsilon_i(\mathbf{p})$ are
$MN$ functions to be determined. The latter correspond to the eigenvalues
$\pm
\epsilon_i(\mathbf{p})$ of the matrix $A(\mathbf{p})$.
Their exact form has to be calculated separately for each choice of unit
cell and vortex configuration. The ground state and the first excited state
corresponding to a particular vortex configuration are given by
\begin{eqnarray}
    \ket{gs} = \prod_{i=1}^{MN} \prod_{-\pi \leq p_x, p_y \leq \pi} b_i(\mathbf{p}) \ket{0}, \qquad \ket{1} = b_1^{\dagger}(\mathbf{p}_0) \ket{gs},
\end{eqnarray}
where $\ket{0}$ is a state with no Majorana fermions and $\mathbf{p}_0$ is the momentum minimizing the lowest lying eigenvalue $\epsilon_1(\mathbf{p})$, i.e. $\min_{\mathbf{p}} |\epsilon_1(\mathbf{p})| = \epsilon_1(\mathbf{p}_0)$. It follows that the corresponding total ground state energy, $E$, and the fermion gap, $\Delta$, are given by
\begin{eqnarray}
    E & = & - \int_{-\pi/M}^{\pi/M} \frac{d p_x}{2\pi/M} \int_{-\pi/N}^{\pi/N} \frac{d p_y}{2\pi/N} \sum_{i=1}^{MN} \frac{|\epsilon_i(\mathbf{p})|}{2}, \label{Egs} \\
    \Delta & = & \min_{\mathbf{p}} |\epsilon_1(\mathbf{p})|. \label{fermiongap}
\end{eqnarray}

\section{Analytic results at the thermodynamic limit}
\label{Analytic}

In this section we present analytic solutions to the two limiting vortex
configurations: the vortex-free and full-vortex configurations. Furthermore,
we outline how the generalized unit cells can be used to study
configurations, where the separation between two vortices is varied. This
will be later used to study the behavior of the relative ground state
energies and fermion gaps as the function of the vortex separation.

\subsection{The vortex-free configuration}

\begin{figure}[t]
\begin{center}
\begin{tabular}{cc}
\includegraphics*[width=6cm]{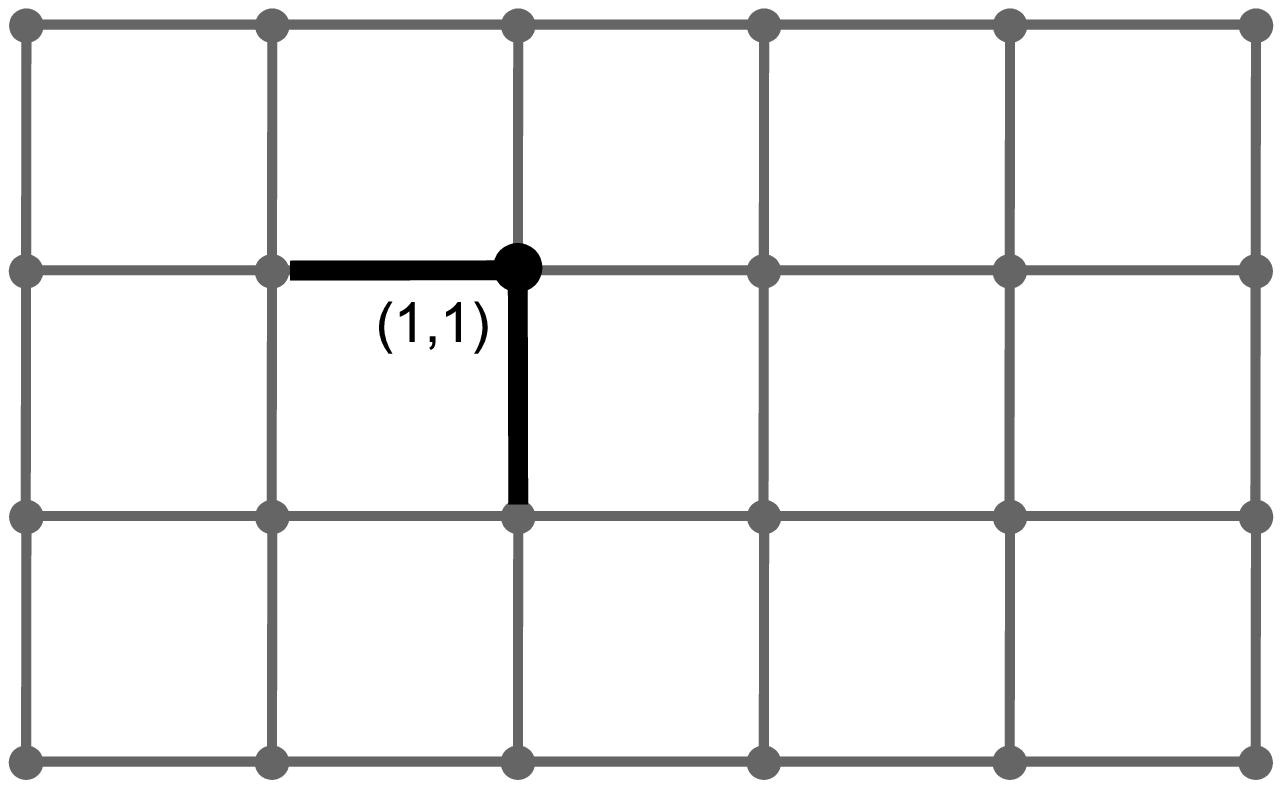} &
\includegraphics*[width=6cm]{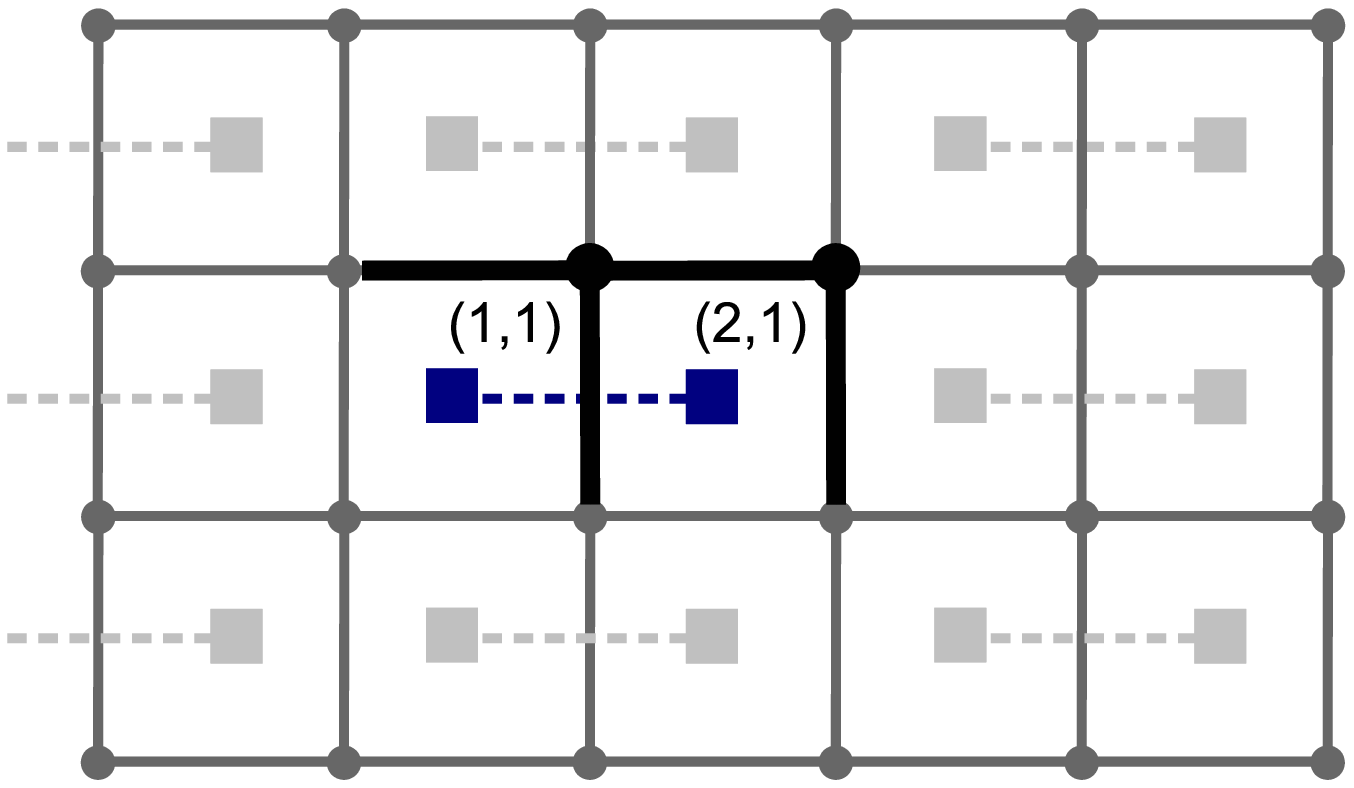}
 \\ (a) & (b) \end{tabular} \end{center}
\caption{\small{(a) The vortex-free configuration is created by setting $u=1$ on all links. The configuration is periodic with respect to a 
$(1,1)$-unit cell. (b) The
full-vortex configuration is created by alternating the value of the $u$'s on $y$-links in the $x$-direction. We take $u=1$ on all $x$- and $z$-links. The configuration is periodic with respect to a $(2,1)$-unit cell. The solid squares denote the location of the vortices and the dashed lines indicate the strings along which $u=-1$.}}
\label{uc_11_21}
\end{figure}

The vortex-free configuration is achieved by setting
\begin{eqnarray}
    u_{k,l} =  1, \forall k,l.
\end{eqnarray}
This configuration is periodic with respect to each $z$-link and thus we can
choose a $(1,1)$-unit cell (see Figure~\ref{uc_11_21}(a)). The off-diagonal,
\rf{A12}, and diagonal,~\rf{A11}, elements are then given by
\begin{eqnarray}
    A_{12}(\mathbf{p}) & = & 2i \left( J_z + J_x e^{i\mathbf{p}\cdot \mathbf{v}_x} + J_y e^{i\mathbf{p} \cdot \mathbf{v}_y} \right) = if(\mathbf{p}), \nonumber \\
    A_{11}(\mathbf{p}) & = & 4K \left(\sin[\mathbf{p} \cdot (\mathbf{v}_x-\mathbf{v}_y) ] + \sin(\mathbf{p} \cdot \mathbf{v}_y) - \sin(\mathbf{p} \cdot \mathbf{v}_x) \right) = g(\mathbf{p}). \nonumber
\end{eqnarray}
Inserting these together with \rf{A12rel} into \rf{H2}, we obtain a $2 \times 2$ Hamiltonian which is diagonalized by introducing the fermionic operator
\begin{eqnarray}
  b(\mathbf{p}) = \Lambda \left( c_2(\mathbf{p}) +
  i \frac{\epsilon(\mathbf{p}) -g(\mathbf{p})}{f(\mathbf{p})}
  c_1 (\mathbf{p})\right),
  \Lambda^2 = \frac{|f(\mathbf{p})|^2}{(\epsilon(\mathbf{p}) +
  g(\mathbf{p}))^2 + |f(\mathbf{p})|^2},
  \nonumber
\end{eqnarray}
where
\begin{eqnarray}
        \epsilon(\mathbf{p}) & = & \sqrt{|f(\mathbf{p})|^2+g(\mathbf{p})^2}, \nonumber \\
    |f(\mathbf{p})|^2 & = & 4(J_x^2 + J_y^2 + J_z^2 + 2\left( J_x J_z \cos p_x + J_x J_y \cos (p_x-p_y) + J_y J_z \cos p_y \right)), \nonumber\\
    g(\mathbf{p})^2 & = & 16K^2 \left(\sin p_y -\sin p_x + \sin (p_x-p_y) \right)^2. \nonumber
\end{eqnarray}
The eigenvalues of the Hamiltonian are given by $\pm \epsilon(\mathbf{p})$,
and thus the total ground state energy \rf{Egs} and the fermion gap
\rf{fermiongap} are given by
\begin{eqnarray}
    E_{0} & = & - \int_{-\pi}^{\pi} \frac{dp_x}{2\pi} \int_{-\pi}^{\pi} \frac{dp_y}{2\pi} \ \frac{\epsilon(\mathbf{p})}{2}, \label{E0} \\
    \Delta_{0} & = & \min_{\mathbf{p}} |\epsilon(\mathbf{p})|. \label{vfgap}
\end{eqnarray}
These results agree with the ones obtained in~\cite{Kitaev05} and~\cite{Pachos06}.

\subsection{The full-vortex configuration}

The full-vortex configuration can be obtained by choosing a
$(2,1)$-unit cell and setting
\begin{eqnarray}
u_{k,l} = \left\{ \begin{array}{l} -1, \quad k =(1,1) \ \textrm{and} \ l =
k-n_y, \\ 1, \quad \quad \textrm{otherwise}. \end{array} \right. .
\end{eqnarray}
Figure \ref{uc_11_21}(b) illustrates the choice of unit cell for this case.
Equations~\rf{A12} and \rf{A11} become
\begin{eqnarray}
    A_{12} = 2i \left( \begin{array}{cc} J_z - J_y e^{i\mathbf{p} \cdot \mathbf{v}_y} & J_x e^{i\mathbf{p} \cdot \mathbf{v}_x } \\ J_x & J_z + J_y e^{i\mathbf{p} \cdot \mathbf{v}_y} \end{array}  \right) \nonumber
\end{eqnarray}
and
\begin{eqnarray}
    A_{11} = 2iK \left( \begin{array}{cc} e^{i\mathbf{p} \cdot \mathbf{v}_y} - e^{-i\mathbf{p} \cdot \mathbf{v}_y} & e^{i\mathbf{p} \cdot \mathbf{v}_x} - 1 - e^{i\mathbf{p} \cdot \mathbf{v}_y} - e^{i\mathbf{p} \cdot ( \mathbf{v}_x- \mathbf{v}_y)} \\ e^{i\mathbf{p} \cdot (\mathbf{v}_y - \mathbf{v}_x)} - e^{-i\mathbf{p} \cdot \mathbf{v}_x} + 1 + e^{-i\mathbf{p} \cdot \mathbf{v}_y} & -e^{i\mathbf{p} \cdot \mathbf{v}_y} + e^{-i\mathbf{p} \cdot \mathbf{v}_y} \end{array}  \right). \nonumber
\end{eqnarray}
Inserting these into \rf{H2} and diagonalizing the resulting $4 \times 4$ Hamiltonian we
obtain the eigenvalues
\begin{eqnarray} \label{eigenvalues}
    \epsilon (\mathbf{p}) = \pm 2 \sqrt{f(\mathbf{p}) \pm 2\sqrt{g(\mathbf{p})}},
\end{eqnarray}
where
\begin{eqnarray}
    f(\mathbf{p}) & = & J_x^2+J_y^2+J_z^2 + 4K^2(\sin^2(p_x-p_y)+\sin^2 p_y + \cos^2 p_x), \nonumber \\
    g(\mathbf{p}) & = & J_x^2 J_y^2 \cos^2(p_x-p_y) + J_x^2 J_z^2 \sin^2 p_x + J_y^2 J_z^2 \cos^2 p_y +  \nonumber \\
    \ & \ & 4K^2 \Big[ J_x^2 \sin^2 p_y + J_y^2 \cos^2 p_x + J_x^2 \sin^2(p_x-p_y) \nonumber \\
    \ & \ & \qquad \qquad  - (J_x J_y + J_x J_z + J_y J_z)\sin(p_x-p_y)\sin p_y \cos p_x \Big]. \nonumber
\end{eqnarray}
The analytic expressions of the corresponding eigenvectors are not presented here as they are too lengthy. When we set $K=0$ our results agree with the analytic
calculations performed in \cite{Pachos06} in the absence of the $K$-term. The
total ground state energy \rf{Egs} is now given by
\begin{eqnarray} \label{Efv}
    E_{fv} & = & - \int_{-\pi/2}^{\pi/2} \frac{dp_x}{\pi} \int_{-\pi}^{\pi} \frac{dp_y}{2\pi} \left( \sqrt{f(\mathbf{p}) + 2\sqrt{g(\mathbf{p})}} + \sqrt{f(\mathbf{p}) - 2\sqrt{g(\mathbf{p})}} \right),
\end{eqnarray}
and the fermionic gap \rf{fermiongap} becomes
\begin{eqnarray} \label{vlgap}
    \Delta_{fv} = \min_{\mathbf{p}} \left| 2\sqrt{f(\mathbf{p}) - 2\sqrt{g(\mathbf{p})}} \right|.
\end{eqnarray}

\subsection{Sparse vortex configurations}

We turn next to sparse vortex configurations in order to study the interactions between vortices. This is done by considering how the ground
state energies and fermion gaps of 2-vortex configurations behave when the separation between vortices is varied. A configuration with two vortices separated by
$s$ plaquettes is created, for instance, by selecting an $(M,N)$-unit cell
and setting
\begin{eqnarray} \label{u_sparse}
    u_{k,l} = \left\{ \begin{array}{ll} -1, &  \qquad k = (1 \leq m \leq s,1) \ \textrm{and} \ l = k-n_y \\ 1, & \qquad \textrm{otherwise}.  \end{array} \right.
\end{eqnarray}
Assuming $M \geq N$, vortex separations of $s < M/2$ can be studied.
\begin{figure}[t]
\begin{center}
\includegraphics*[width=8cm]{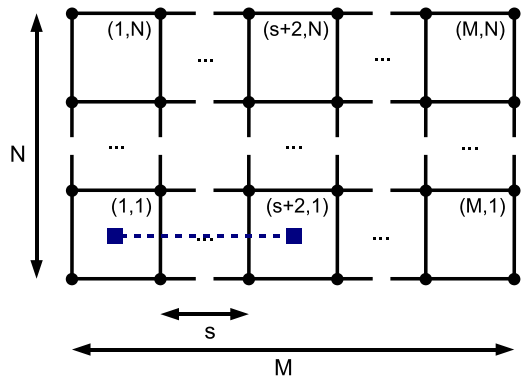}
\end{center}
\caption{\label{sparsecell}\small{$(M,N)$-unit cell containing a pair
of vortices separated by $s$ plaquettes. $s=0$ means that the vortices occupy neighboring plaquettes. The solid squares in the plaquettes
indicate the location of the vortices and the dashed line indicate the
string along which $u=-1$ on all $y$-links. On all other links $u=1$.}}
\end{figure}
Figure \ref{sparsecell} illustrates the unit cell and the resulting vortex configuration.
Since we are working on an infinite plane, ideally one would like to use a
unit cell of infinite size in order to isolate the interaction between the vortices. However, the Hamiltonian \rf{H2} grows polynomially in $M$ and
$N$ and hence we restrict to considering a $(20,20)$-unit cell. This choice
allows separations $s\leq9$, which is shown later to be sufficient to
extract the expected asymptotic behavior when $s \to \infty$. The resulting
Hamiltonians are sparse $800 \times 800$ matrices, which can not be treated
analytically, but can be diagonalized numerically in a reasonable time using
a tabletop computer. Using \rf{Egs} and \rf{fermiongap} we can then
calculate the total ground state energies $E^s_{2v}$ and fermion gaps
$\Delta_{2v}^s$ corresponding to 2-vortex configurations with vortex
separation $s$.

\section{Analysis of the spectrum}
\label{analysis}

The spectrum of Hamiltonian (\ref{H}) can be characterized by two different
types of energy gaps: fermionic gaps that characterize the energy levels of the
spectrum above a fixed vortex configuration and vortex gaps that
compare the ground state energies corresponding to different vortex configurations. We determine how
the presence of vortices influences the fermionic gaps and, subsequently,
how the phase space geometry is modified. We also determine the scaling of the ground state degeneracy, which is expected due to the presence of the Ising
non-abelian vortices. Moreover, we carry out a study on the 2-vortex configuration energies as a function of the vortex separation, and subsequently determine the vortex gap, i.e. the energy required to
excite a pair of free vortices.

\subsection{The fermion gap}

\subsubsection{The phase space geometry in the presence of vortices}

\begin{figure}[p]
\begin{center}
\includegraphics*[width=7cm]{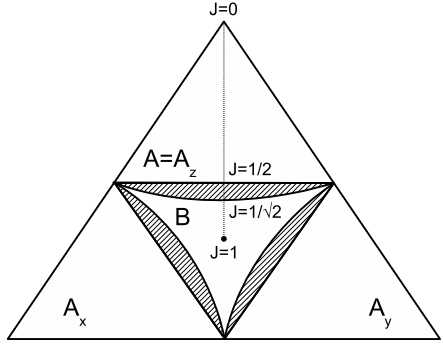}
\end{center}
\caption{\label{phasediagram} \small{ 
An illustration of the phase diagram with the four distinct phases $A_x, A_y, A_z$ and $B$ when $J_x+J_y+J_z=1$. We restrict to studying the transition only between $A \equiv A_z$ and $B$ phases and for convenience employ an alternative normalization such that $J = J_x = J_y$ and $J_z = 1$.  The dashed line indicates the $0 \leq J \leq 1$ part of the phase space
along which we study the system. For all vortex configurations at small $K$ the phase boundary between $A$ and $B$ phases falls into the shaded area in the
inner triangle. The limiting phase boundaries are given by the vortex-free $(J=1/2)$ and full-vortex $(J=1/\sqrt{2})$ configurations.}}
\end{figure}

First, we briefly review the phase space of the vortex-free sector, which
was studied in~\cite{Kitaev05}. It was shown that the honeycomb lattice
model exhibits four distinct phases $A_x, A_y, A_z$ and $B$ for different
values of the couplings $J_\alpha$ such that the system is in the $B$-phase
when all the inequalities $|J_y|+|J_z| \leq |J_x|$, $|J_x|+|J_z| \leq |J_y|$
and $|J_x|+|J_y| \leq |J_z|$ are violated. The phase boundaries are given by
the equalities and the phase $A_\alpha$ occurs when only
$|J_\beta|+|J_\gamma|
\leq |J_\alpha|$ holds and the other two inequalities are violated. The
$A_\alpha$ phases are always gapped for $J_\beta,J_\gamma\neq0$, $K\geq 0$
and the vortices behave as $Z_2 \times Z_2$ abelian anyons. On the other
hand, the $B$-phase is gapped only when $K
\neq 0$ and only there the vortices behave as non-abelian Ising anyons.
The phase boundaries are the lines in the phase space where the fermion gap
vanishes. Here we restrict to studying the transition, i.e. the behavior of
the fermion gap between the $A=A_z$ (abelian) and the $B$ (non-abelian)
phases. Figure~\ref{phasediagram} illustrates the
general phase space geometry where we've taken $J_x+J_y+J_z=1$. For convenience we normalize from now on the couplings such
that $J_z=1$ and $J_x = J_y = J$.

\begin{figure}[p]
\begin{center}
\begin{tabular}{cc}
\includegraphics*[width=6.5cm]{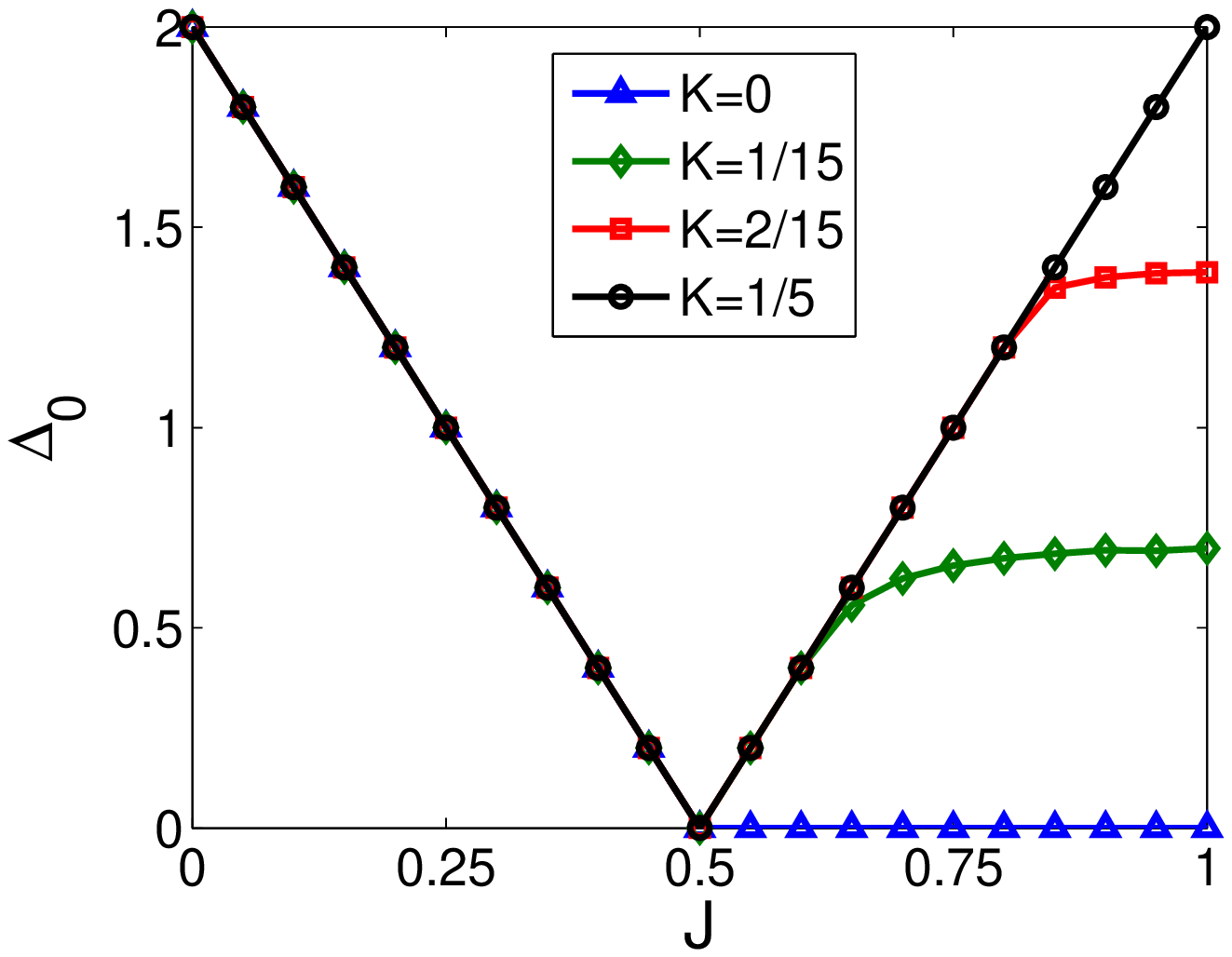} & \includegraphics*[width=6.5cm]
{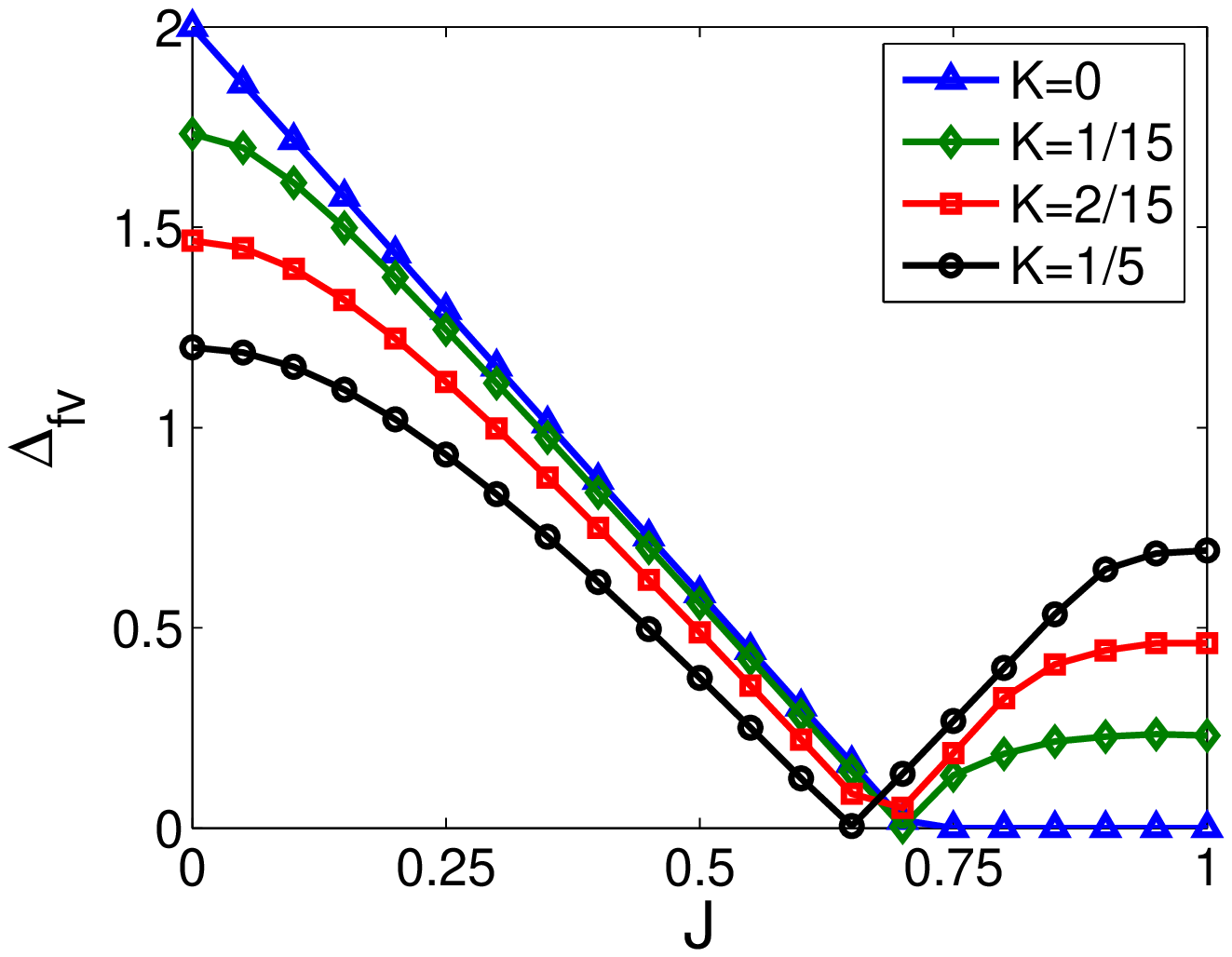} \\ (a) & (b)
\end{tabular} \end{center}
\caption{ \small{(a) The vortex-free, \rf{vfgap}, and (b) the full-vortex, \rf{vlgap}, configuration fermion gaps for different values of $K$. In the vortex-free case the gap vanishes at $J=1/2$ for all values of $K$. In the full-vortex case the gap vanishes at $J=1/\sqrt{2}$ for $K=0$, but shifts to smaller $J$ as $K$ is increased.}}
\label{vlvfgapplot}
\end{figure}

Figures \ref{vlvfgapplot}(a) and \ref{vlvfgapplot}(b) show the
vortex-free,~\rf{vfgap}, and full-vortex,~\rf{vlgap}, fermionic gaps plotted
as functions of $J$ for different values of $K$. Let us first consider the
$K=0$ case. In the vortex-free configuration the gap vanishes at $J=1/2$, in
agreement with~\cite{Kitaev05}. In the full-vortex configuration the gap
persists till $J=1/\sqrt{2}$ in line with the results derived
in~\cite{Pachos06}. There the phase boundary equalities for the full-vortex
configuration were derived to take the form $|J_\beta|^2+|J_\gamma|^2 =
|J_\alpha|^2$. When $K\neq 0$, in both cases the fermion gaps reappear and
settle at a constant value once the system moves well into the $B$ phase.
However, the dependence of the gap magnitude on $K$ is clearly different for
the vortex-free and full-vortex configurations with the gap being much
smaller in the latter case. Also, Figure \ref{vlvfgapplot}(b) shows that at $K=1/5$ the transition between the phases is shifted away from $J=1/\sqrt{2}$. This implies that the magnitude of $K$ can also affect the phase space geometry. 

In the limiting cases we observe that the boundary between the two phases
depends on the underlying vortex configuration. To study how the fermionic
energy gap interpolates in between these two extreme cases, we consider the fermion gaps of various sparse vortex configurations on small $(MN \leq 12)$ unit cells. In all the cases the phase boundary falls into the region $1/2 \leq J \leq 1/\sqrt{2}$ when $K=0$. For very sparse configurations with low vortex density such as the 2-vortex configuration created by \rf{u_sparse}, the boundary is located very close to $J=1/2$, whereas for more homogeneously distributed configurations with larger vortex density it tends towards $J=1/\sqrt{2}$. However, when $K > 0$ the phase boundary is in general shifted to larger $J$'s such that for some configurations the boundary is located in the area $J>1/\sqrt{2}$. We attribute the shifting of the phase boundary to short-range vortex-vortex interactions, which are enhanced when $K$ is increased. It is interesting to note that since the vortex density affects the phase space geometry in the region $1/2 \leq J \leq 1/\sqrt{2}$ and $K \geq 0$, it could, in principle, be used as a tunable parameter, that induces a phase transition between the abelian and non-abelian phase.

\subsection{The fermion gap in the presence of vortices}

\begin{figure}[p]
\begin{center}
\begin{tabular}{cc}
\includegraphics*[width=6.5cm]{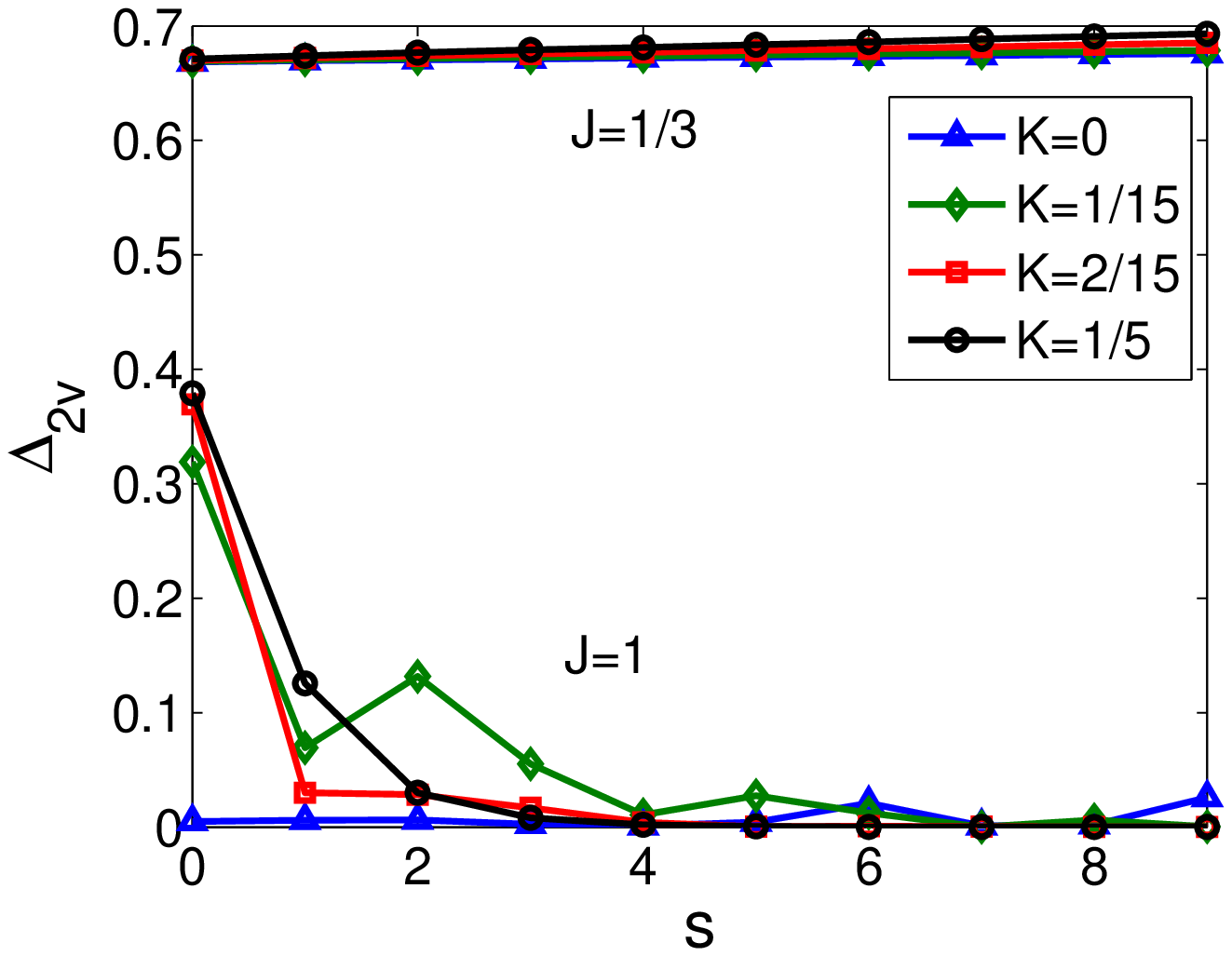} &
\includegraphics*[width=6.5cm]{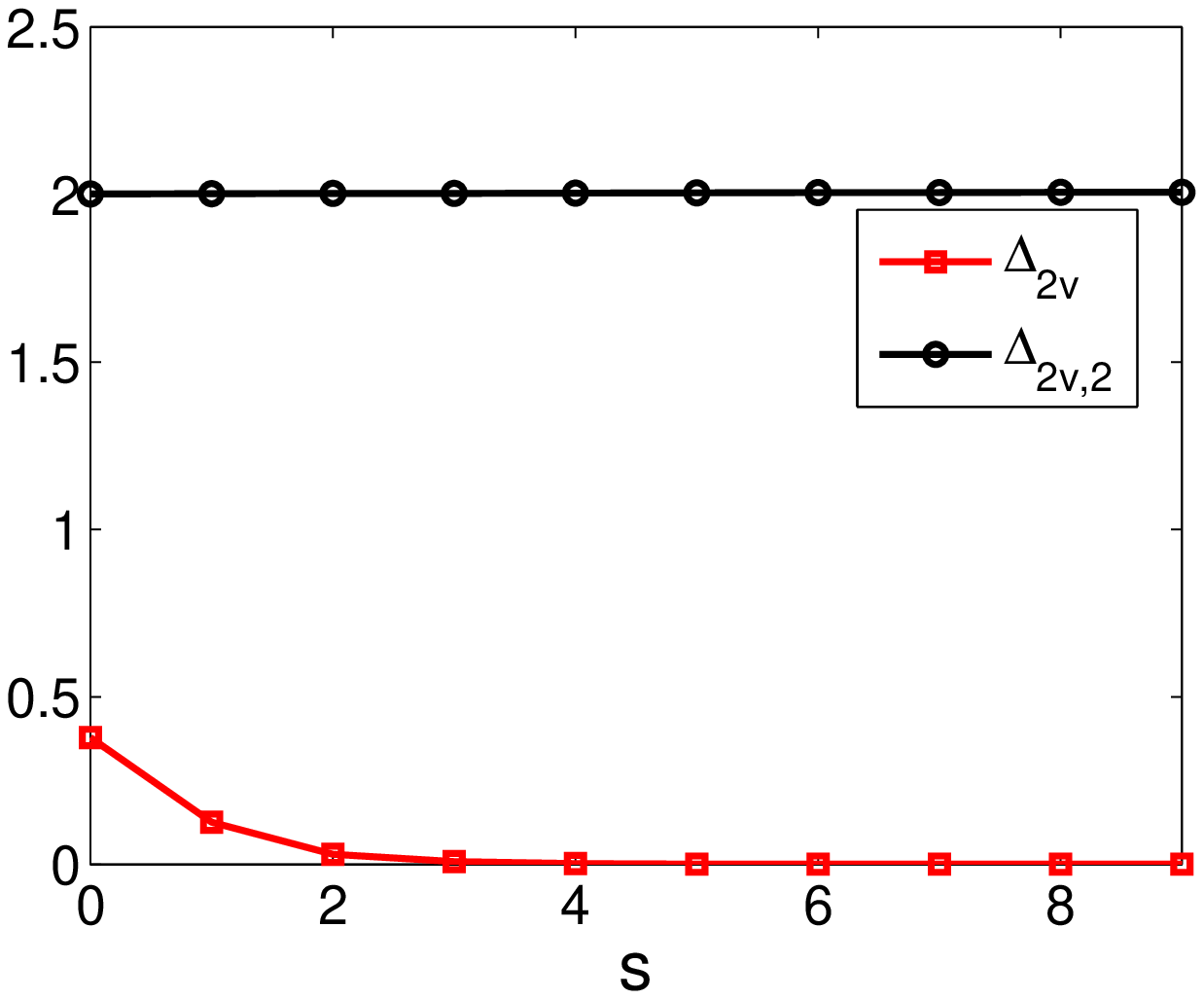}
 \\ (a) & (b)
\end{tabular} \end{center}
\caption{The gap behavior for a 2-vortex configuration as a function of the vortex separation $s$. (a) The fermion gap $\Delta_{2v}$ at the abelian $(J=1/3)$ and non-abelian $(J=1)$ phases for several
values of $K$. (b) The two lowest lying excited states \rf{2vgaps} in the non-abelian phase ($J=1$ and $K=1/5$).
In the non-abelian phase the ground state is a twofold degenerate for
well separated vortices. The degeneracy is lifted when the two vortices are brought close
to each other.}
\label{fgaps_2v}
\end{figure}

It is intriguing to study the behavior of the fermionic gap in the presence
of only two vortices as a function of their distance. For that we consider again
the 2-vortex configurations with varying vortex separation $s$, which are created by setting the $u$'s as given by~\rf{u_sparse}. Figure~\ref{fgaps_2v}(a) shows the behavior of the corresponding fermion
gap $\Delta_{2v}$ as the function of $s$ in the abelian
$(J=1/3)$ and in the non-abelian $(J=1)$ phase for several values of $K$.
The behavior in the different phases is radically different. In the abelian
phase the fermion gap is in practice insensitive to both $s$ and $K$. In stark contrast to the abelian phase, the fermion gap in the non-abelian phase decreases exponentially with $s$ vanishing completely for $s>2$.\footnote{The oscillations of the energy gap as a function of the vortex separation appear to be Friedel oscillations.} Indeed, in
Figure~\ref{fgaps_2v}(b) we plot the gaps to the two first excited states,
\begin{eqnarray} \label{2vgaps}
\begin{array}{rclcrcl}
	\ket{1} & = & b^{\dagger}_1 (\mathbf{p}_0) \ket{gs}, & \qquad & \Delta_{2v} & = & \min_{p_0} |\epsilon_1(\mathbf{p})|, \\
	\ket{2} & = & b^{\dagger}_2 (\mathbf{p}_0) \ket{gs}, & \qquad & \Delta_{2v,2} & = & \min_{p_0} |\epsilon_2(\mathbf{p})|,
\end{array}
\end{eqnarray}
and observe that $\Delta_{2v}$ tends exponentially to zero as $s$ increases with the value at $s=9$ being of order $10^{-7}$. This means that in the
presence of two well separated vortices the ground state of the non-abelian phase is twofold
degenerate. Moreover, the gap to the second excited state, $\Delta_{2v,2}$, is found to be insensitive to $s$ and persist to arbitrary separations. 

\begin{figure}[p]
\begin{center}
\begin{tabular}{cc}
\includegraphics*[width=6.5cm]{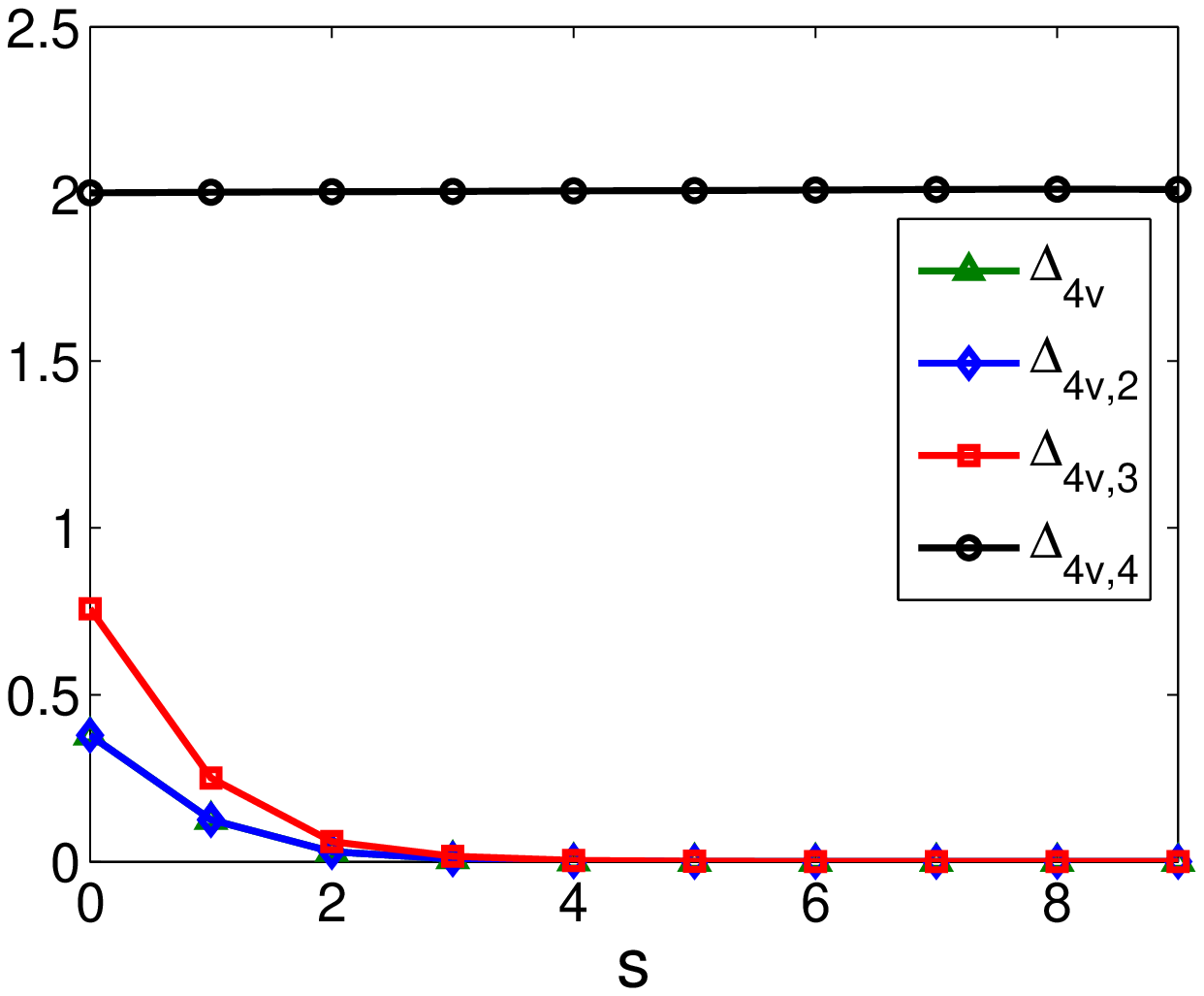} &
\includegraphics*[width=6.5cm]{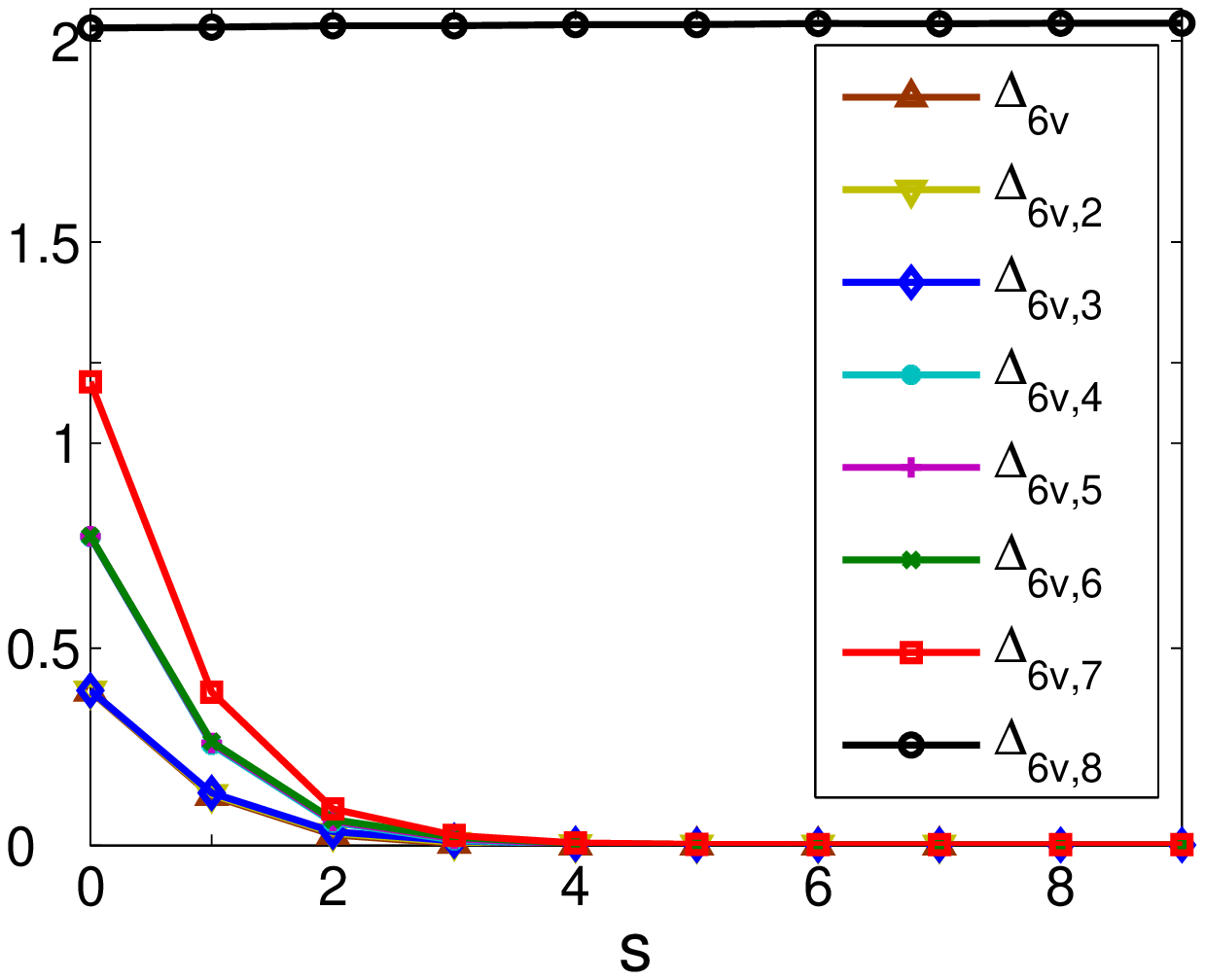} \\ (a) & (b)
\end{tabular} \end{center}
\caption{\small{ The fermion gaps
for some of the first excited states at $J=1$ and $K=1/5$ as a function of
the vortex separation $s$. (a) The fermion gaps \rf{4vgaps} to the four first excited sates of the 4-vortex
configuration. The first and second excited states are degenerate. (b)
The fermion gaps \rf{6vgaps} of the eight first excited states for the 6-vortex configuration. 1st, 2nd and 3rd as well as 4th, 5th and 6th excited states remain degenerate at small $s$.}}
\label{fgaps}
\end{figure}

It is known that the honeycomb lattice model can be mapped to a $p$-wave
superconductor where the fermions live on the $z$-links \cite{Chen,Yu}. The
vortices in the superconductor are assumed to be non-abelian Ising
anyons \cite{Read}, which are also predicted to appear as vortices in the
honeycomb lattice model \cite{Kitaev05}. In the presence of $2n$ well
separated vortices the ground state should be $2^n$-fold degenerate,
but when vortices are brought together, their interactions are predicted to
lift the degeneracy \cite{Read,Gurarie,Stone}. Our demonstration of the
twofold degenerate ground state in the presence of two vortices is in
agreement with this prediction.  To verify that
the degeneracy scales as $2^n$ for the honeycomb lattice model, we consider
in addition 4- and 6-vortex pairwise configurations where the vortex pairs are
located on equally spaced rows of the $(20,20)$-unit cell and the separation $s$ of the vortices from each pair is simultaneously varied. In the 4-vortex case the fermion gaps to the four first excited states are given by
\begin{eqnarray} \label{4vgaps}
\begin{array}{rclcrcl}
	\ket{1} & = & b^{\dagger}_1 (\mathbf{p}_0) \ket{gs}, & \qquad & \Delta_{4v} & = & \min_{p_0} |\epsilon_1(\mathbf{p})|, \\
	\ket{2} & = & b^{\dagger}_2 (\mathbf{p}_0) \ket{gs}, & \qquad & \Delta_{4v,2} & = & \min_{p_0} |\epsilon_2(\mathbf{p})|, \\
	\ket{3} & = & b^{\dagger}_1 (\mathbf{p}_0) b^{\dagger}_2 (\mathbf{p}_0) \ket{gs}, & \qquad & \Delta_{4v,3} & = & \min_{p_0} |\epsilon_1(\mathbf{p}) + \epsilon_2(\mathbf{p})|, \\
	\ket{4} & = & b^{\dagger}_3 (\mathbf{p}_0) \ket{gs}, & \qquad & \Delta_{4v,4} & = & \min_{p_0} |\epsilon_3(\mathbf{p})|.
\end{array}
\end{eqnarray}
Similarly, the gaps to the eight first excited states for the 6-vortex configuration are 
\begin{eqnarray} \label{6vgaps}
\begin{array}{rclcrcl}
	\ket{1} & = & b^{\dagger}_1 (\mathbf{p}_0) \ket{gs}, & \qquad & \Delta_{6v} & = & \min_{p_0} |\epsilon_1(\mathbf{p})|, \\
	\ket{2} & = & b^{\dagger}_2 (\mathbf{p}_0) \ket{gs}, & \qquad & \Delta_{6v,2} & = & \min_{p_0} |\epsilon_2(\mathbf{p})|, \\
	\ket{3} & = & b^{\dagger}_3 (\mathbf{p}_0) \ket{gs}, & \qquad & \Delta_{6v,3} & = & \min_{p_0} |\epsilon_3(\mathbf{p})|, \\
	\ket{4} & = & b^{\dagger}_1 (\mathbf{p}_0) b^{\dagger}_2 (\mathbf{p}_0) \ket{gs}, & \qquad & \Delta_{6v,4} & = & \min_{p_0} |\epsilon_1(\mathbf{p}) + \epsilon_2(\mathbf{p})|, \\
	\ket{5} & = & b^{\dagger}_1 (\mathbf{p}_0) b^{\dagger}_3 (\mathbf{p}_0) \ket{gs}, & \qquad & \Delta_{6v,5} & = & \min_{p_0} |\epsilon_1(\mathbf{p}) + \epsilon_3(\mathbf{p})|, \\
	\ket{6} & = & b^{\dagger}_2 (\mathbf{p}_0) b^{\dagger}_3 (\mathbf{p}_0) \ket{gs}, & \qquad & \Delta_{6v,6} & = & \min_{p_0} |\epsilon_2(\mathbf{p}) + \epsilon_3(\mathbf{p})|, \\
	\ket{7} & = & b^{\dagger}_1 (\mathbf{p}_0) b^{\dagger}_2 (\mathbf{p}_0) b^{\dagger}_3 (\mathbf{p}_0) \ket{gs}, & \qquad & \Delta_{6v,7} & = & \min_{p_0} |\epsilon_1(\mathbf{p}) + \epsilon_2(\mathbf{p}) + \epsilon_3(\mathbf{p})|, \\
	\ket{8} & = & b^{\dagger}_4 (\mathbf{p}_0) \ket{gs}, & \qquad & \Delta_{6v,8} & = & \min_{p_0} |\epsilon_4(\mathbf{p})|.
\end{array}
\end{eqnarray}
Figures~\ref{fgaps}(a) and (b) depict the behavior of
the 4- and 6-vortex configuration fermion gaps \rf{4vgaps} and \rf{6vgaps}, respectively, at $J=1$ and $K=1/5$ as a function of $s$. 
The degeneracy of the ground state as $s \to\infty$ is four for 4-vortex and eight for
6-vortex configurations. This is exactly the predicted scaling. We also observe that at $s<2$ the interaction does not completely lift the
degeneracies. The ground state becomes non-degenerate for small $s$, but
some of the states form degenerate bands. The
splitting in energy between the bands is homogenous in the sense that for a specific $s$ it costs the same amount of energy to move between the shifted states. We also observe that for all the considered vortex configurations the first non-vanishing gap as $s \to \infty$ is always two units of energy above the ground state. As can be seen from Figure \ref{vlvfgapplot}(a), this is the case also for the vortex-free fermion gap, $\Delta_0$, at $J=1$ and $K=1/5$. The observation $\Delta_0 = \Delta_{2v,2} = \Delta_{4v,4} = \Delta_{6v,8}$
suggests that the energy to excite a fermionic mode is insensitive to the underlying vortex configuration and depends only on $J$ and $K$. We will adopt $\Delta_0$ to denote the energy to create a \emph{free} fermionic excitation as opposed to an excited state at small $s$ due to lifting of the ground state degeneracy.

\subsubsection{Fermionic spectrum and the Ising anyon model}

Our results concerning the spectrum can be interpreted in the context of
Ising anyon model, which is assumed to describe the low energy behavior
of both the honeycomb lattice model and the $p$-wave paired superconductors
\cite{Read,Kitaev05}. This model is spanned by three types of
quasiparticles or sectors: the vacuum $1$, a fermion $\psi$ and a
non-abelian $\sigma$. The non-trivial fusion rules are given by
\begin{eqnarray} \label{fusion}
    \psi \times \psi = 1, \quad \psi \times \sigma = \sigma, \quad  \sigma \times \sigma = 1 + \psi.
\end{eqnarray}
In the context of $p$-wave paired superconductors, $1$ can be understood as
the ground state condensate of Cooper pairs, $\psi$ as a Bogoliubov
quasiparticle and $\sigma$ as a vortex \cite{Stone}. In the honeycomb
lattice model we take analogously $1$ to be the ground state, $\psi$ to be a fermion mode $b_i$ in the spectrum \rf{H_diag} and $\sigma$ to
be a vortex living on a plaquette. 

An established method to study $p$-wave superconductor vortices is in terms
of massless Majorana modes $\gamma_i$ localized inside the vortex cores
\cite{Read,Ivanov,Stern,Stone}. Two Majorana modes can be combined to a
fermion mode $z_i = (\gamma_i+i\gamma_{i+1})/\sqrt{2}$, which is
carried by a pair of vortices located at $i$ and $i+1$. Whether this mode is
occupied or unoccupied corresponds to the two possible fusion outcomes of
the $\sigma$ vortices - unoccupied mode corresponds to fusing to vacuum $1$
whereas occupied mode means that the fusion will yield a $\psi$. Since the occupation of these modes does not increase the energy of the system, they are known as zero modes, which appear in the spectra of systems supporting Ising anyons. The existence of $n$ zero modes in the spectrum implies $2^n$-fold
degenerate ground state. However, when the vortices are brought close to each
other, the degeneracy should be lifted in a way that allows the determination of the fusion outcome
\cite{Kitaev, Read, Gurarie}.

The observed degeneracy in the presence of well separated vortices in the honeycomb lattice model (see Figures \ref{fgaps_2v}, \ref{fgaps}(a) and \ref{fgaps}(b)) can be explained in terms of the ``zero energy
modes" in the spectrum. These modes do not strictly speaking have zero energy, but correspond instead to modes with the same finite energy as the ground state~\rf{Egs}. When $n$ of these zero modes are present, we expect the diagonalized Hamiltonian~\rf{H_diag} to be of the form
\begin{eqnarray} \label{Hzm}
    H & = & MN \int_{-\pi/M}^{\pi/M} \frac{d p_x}{2\pi}
    \int_{-\pi/N}^{\pi/N} \frac{d p_y}{2\pi}
    \left[ \sum_{i=n+1}^{MN} |\epsilon_i(\mathbf{p})|
    b_i^{\dagger} b_i + \sum_{i=1}^{n} |\alpha_i^s(\mathbf{p})|
    z_i^{\dagger} z_i \right. \nonumber \\\ & \ & \left.
    \quad\qquad \qquad \qquad \qquad \qquad  - \left(\sum_{i=n+1}^{MN}
    \frac{|\epsilon_i(\mathbf{p})|}{2} + \sum_{i=1}^{n}
    \frac{|\alpha_i^s(\mathbf{p})|}{2} \right) \right].
\end{eqnarray}
Here $\alpha_i^s(\mathbf{p})$ are the $n$ smallest eigenvalues that vanishes
as the distance between the vortices goes to infinity, i.e.
$\lim_{s\to\infty}\min_{\bf p} |\alpha_i^s(\mathbf{p})| = 0$. We allow $\alpha_i^s(\mathbf{p})$ to be finite at small $s$ to account for the lifting of the ground state degeneracy. This is exactly what we obtain in the presence of vortices. Figures \ref{fgaps_2v}(b), \ref{fgaps}(a) and \ref{fgaps}(b) show that for all $s$ every occupied zero mode contributes equally to the energy splitting, which suggests $\alpha_i (\mathbf{p}) = \alpha (\mathbf{p}), \forall i$. This is reasonable, because occupied zero modes are interpreted as two $\sigma$'s fusing to a $\psi$, and thus every mode at every $s$ should contribute an equal energy proportional to the energy of a $\psi$. 

The splitting of the degenerate ground state in short ranges into degenerate bands spanned by states with the number of occupied zero modes can be used to extract
information about how the vortices fuse. In particular, it is possible to identify the states with different fusion channels. Consider for instance the 4-vortex configuration consisting of two well separated pairs whose fermion gap
behavior is shown in Figure
\ref{fgaps}(b). The fusion rules \rf{fusion} give
\begin{eqnarray}
    \sigma \times \sigma \times \sigma \times \sigma = 1 + 1 + \psi + \psi, \nonumber
\end{eqnarray}
which mean that the four vortices may fuse to both vacuum $1$ and to $\psi$
in two distinct ways. These altogether four distinct fusion channels correspond
to the fourfold degenerate ground state at large $s$. At small $s$ there are
three bands of different energy. The ground state corresponds to fusing both pairs into
vacuum (no occupied zero modes),
\begin{eqnarray}
    (\sigma \times \sigma) \times (\sigma \times \sigma) \to 1
    \times 1 = 1. \nonumber
\end{eqnarray}
On the other hand, the non-degenerate band of two occupied zero modes
corresponds also to the vacuum sector, but now such that both pairs will separately fuse to a $\psi$, 
\begin{eqnarray}
    (\sigma \times \sigma) \times (\sigma \times \sigma) \to
    \psi \times \psi = 1. \nonumber
\end{eqnarray}
Even though this state belongs to the vacuum sector, they differ in energy because the two pairs are well separated from each other.  This contrasts with the twofold degenerate band, which contains the
states corresponding to the two fusion channels 
\begin{eqnarray}
    (\sigma \times \sigma) \times (\sigma \times \sigma) \to
    1 \times  \psi = \psi \quad {\rm and}\quad
(\sigma \times \sigma) \times (\sigma \times \sigma) \to
    \psi \times  1 = \psi. \nonumber
\end{eqnarray}
Both belong to the $\psi$ sector, but there is no energy splitting indicating which pair will fuse to $\psi$ and which to $1$. This is actually only a feature of our construction where the vortices of the $n$ pairs
are always equidistantly separated by $s$ plaquettes. As shown above, it is then only possible to
deduce the total topological sector of all the vortices, and some
information about the global fusion channel by distinguishing between these
states. However, if vortices from
only one pair were brought close to each other, while the others were kept
well separated, the interaction induced gap would correspond to a splitting of
only one zero mode and the first excited state would be non-degenerate. Studying the splitting of each mode separately allows unambiguous determination of the global fusion channel. 

We also comment on the observation that the degree of degeneracy obtained
here is $2^n$. Usually one talks about creating vortices from vacuum, which
means that the global sector of the system is fixed to $1$. The degeneracy corresponding to
$2n$ vortices is then $2^{n-1}$, because when all the vortices are fused,
one must obtain again the vacuum. Our observation of $2^n$-fold degeneracy
means that we do not create vortices out of vacuum by fixing the $u$'s over the unit cell. The overall sector of $2n$ vortices may be either 1 or $\psi$, but we have no prior
information about it. This is reflected also on the identification of the fermion mode $b_i$ with a single $\psi$ excitation. If the overall sector was fixed, the $\psi$'s should always appear in pairs.

\subsection{The vortex gap and the interaction energy}

We have studied the fermionic spectrum above a fixed background vortex
configuration. Here we consider the energy spectrum of the vortices by
studying how the ground state energy depends on the number of vortices
and their separation. A flux phase conjecture proven by Lieb \cite{Lieb}
states that in the absence of an external field the energy minimum for
honeycomb lattice is achieved with a vortex-free configuration. Even though
we have an external magnetic field in our model, we assume this still holds
when $K$ is small. Under this assumption we define the vortex gap
asymptotically by
\begin{eqnarray} \label{vortexgap}
    \Delta E_{2v} = \lim_{M,N \to \infty} (E_{2v}^{M/2} - MN E_{0}),
    \qquad M \geq N,
\end{eqnarray}
which gives the energy to create a pair of vortices and drag them infinitely
far from each other. Here $E_0$ is the vortex-free ground state evaluated on
a single plaquette \rf{E0} and $E_{2v}^{s}$ denotes the total ground state
energy of a vortex configuration on a $(M,N)$-unit cell containing a pair of
vortices separated by $s$ plaquettes. The definition is given for a pair of
vortices due to the constraint~\rf{w} on the plaquette operators that
demands vortices to come in pairs. Including the interaction energy in the
vortex gap definition means that~\rf{vortexgap} provides an estimate of the
stability of the topological phase. To be precise, if the temperature of the
system is well below the vortex gap, $T<<\Delta E_{2v}$, spontaneous creation of
stray vortex pairs will be exponentially suppressed.  

\begin{figure}[t]
\centerline{
\begin{tabular}{cc}
\epsfxsize=7cm \epsfysize=5cm \epsfbox{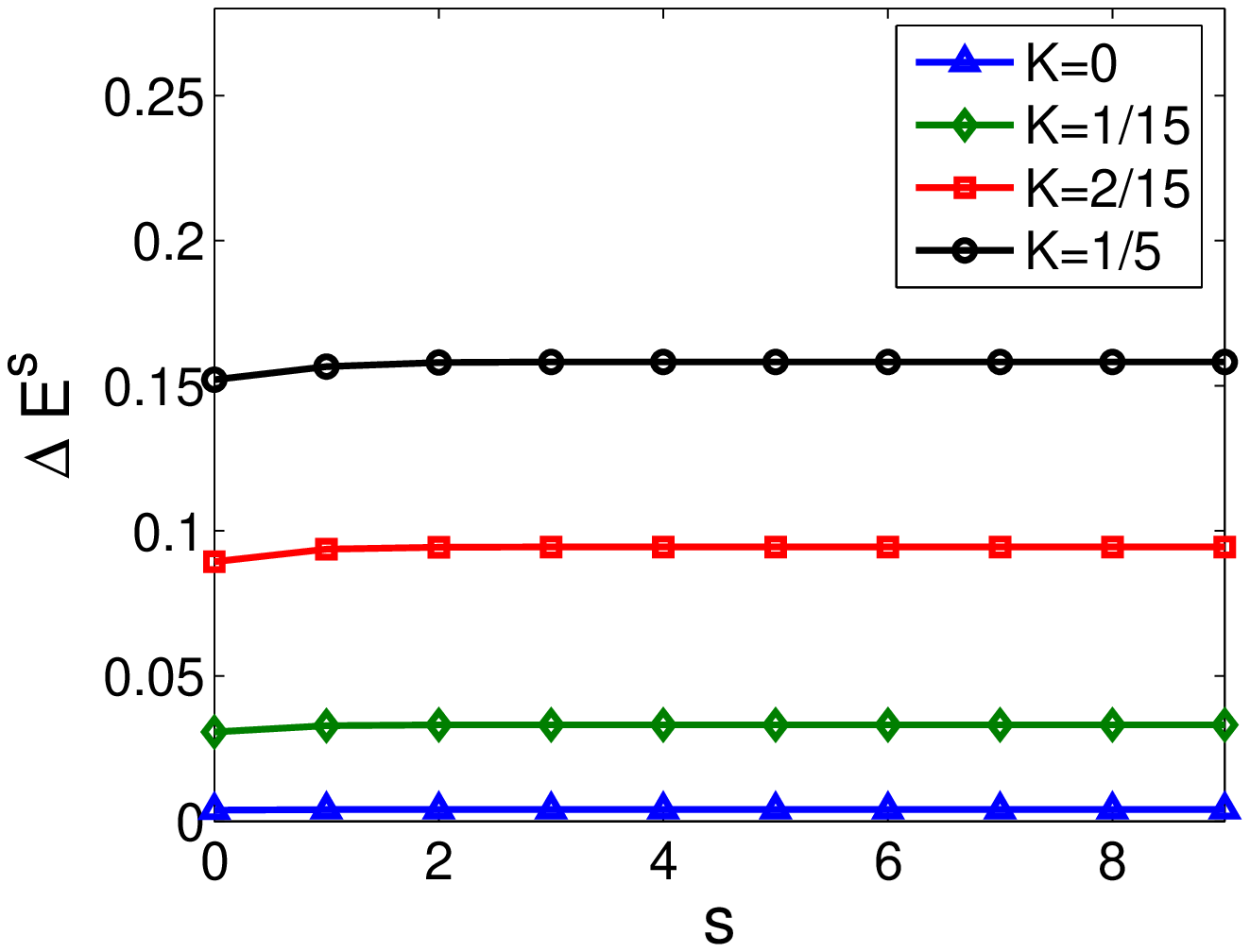} &
\epsfxsize=7cm \epsfysize=5cm \epsfbox{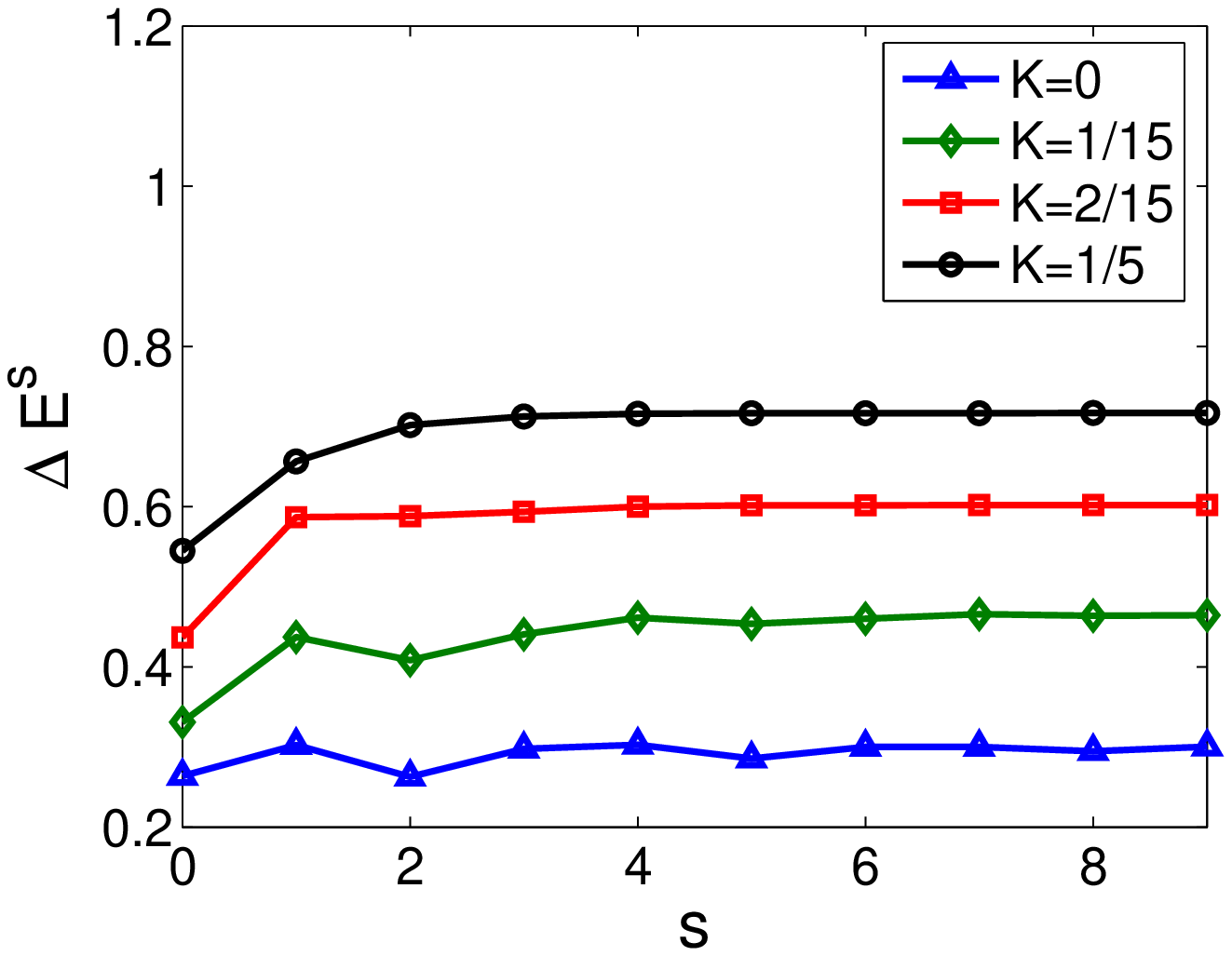} \\
(a) & (b) \end{tabular} }
\caption{\label{homsep} \small{The energy gap $\Delta E^s = E_{2v}^s-20^2 E_0$ between the vortex-free and the 2-vortex configuration with separation $s$ on a $(20,20)$-unit cell in the (a) abelian ($J=1/3$) and in the (b) non-abelian ($J=1$) phase.}}
\end{figure}

To study how this definition applies to systems with periodic
structure, we consider again the $(20,20)$-unit cell with a pair of vortices
separated by $s$ plaquettes~(\ref{u_sparse}). For a particular $s$, the
vortex gap takes the form $\Delta E^s = E_{2v}^s-20^2 E_0$, which is
plotted in Figure \ref{homsep}. The abelian phase (Figure \ref{homsep}(a)) shows a very weak attractive short-range interaction between vortices, which is agreement with the high-order perturbation theory study of the abelian phase \cite{Schmidt}. In the non-abelian phase (Figure \ref{homsep}(b)) we observe also an attractive interaction. However, there the interaction is strong with the magnitude being sensitive to the value of $K$. When $K=0$ the vortex gap exhibits only low amplitude oscillatory
behavior as a function of $s$ void of interaction signature, but as $K$ increases and the system enters the
non-abelian phase, the oscillatory behavior is suppressed and an	
attractive short-range interaction emerges\footnote{Like in Figure \ref{fgaps_2v}(a), the physicality of the oscillatory behaviour for small $K$ is unclear to us.}. In both phases the $K$-term increases the vortex energy, and in the non-abelian phase it is necessary to switch on the interaction. Although the interaction is very weak in the abelian phase, it exists even when $K=0$. We observe that the
interaction becomes negligible when $s > 2$, which is in accordance with the
fermion gap behavior, which was attributed also to the interactions (See Figures~\ref{fgaps_2v}(b), \ref{fgaps}(a) and \ref{fgaps}(b)). 

Our definition contrasts with the vortex gap definition in \cite{Pachos06},
where it was defined as the difference between the total ground state
energies of the full-vortex \rf{Efv} and vortex-free \rf{E0} configurations.
That definition neglected the interaction energy between vortices and hence
we regard our definition \rf{vortexgap} to provide a more accurate estimate
of the energy required to excite the system. It should be emphasized that even though we observe a strong attractive interaction between vortices, the plaquette operators \rf{w} are constants of motion of the Hamiltonian \rf{H} and thus the vortices close to each other are not pulled together and annihilated spontaneously. This is not the case if the $K$-term was replaced by a Zeeman term. Then the vortices could hop and be annihilated if they do not have sufficient energy to overcome the attractive interaction. Therefore, we regard our asymptotic definition of the vortex gap \rf{vortexgap} to provide a realistic way of estimating the stability of the topological phase also in the presence of an external magnetic field.

\subsection{The low energy spectrum of the non-abelian phase}

\begin{figure}[t]
\begin{center}
\includegraphics*[width=8cm]{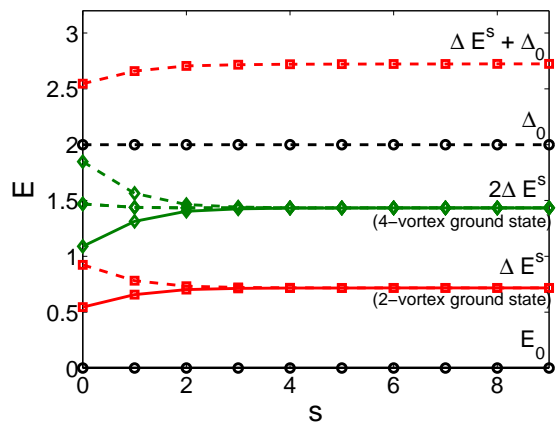}
\end{center}
\caption{\label{lespectrum} \small{
The low energy spectrum of the non-abelian phase ($J=1$ and $K=1/5$) as a function of vortex separation $s$. The plotted energies are with respect to $E_0$, the ground state energy of the vortex-free sector. The solid lines are the total ground state energies and the dashed lines some of the lowest lying excited states above vortex-free (circle), 2- (square) and 4-vortex (diamond) configurations.}}
\end{figure}

Combining the studies on both the fermion and vortex gaps allows us to outline the low-energy spectrum of the non-abelian phase. In Figure \ref{lespectrum} we plot some of the first excited fermionic states above the vortex-free, the 2-vortex and 4-vortex configurations as a function of the vortex separation $s$. All the energies are depicted with respect to the total ground state energy $E_0$ of the vortex-free sector. The vortex-free sector is trivially insensitive to $s$ and the corresponding low energy spectrum is characterized by the fermion gap $\Delta_0$ alone. In terms of the Ising anyon model these fermionic levels were identified with $\psi$ excitations. The 2- and 4-vortex sector ground states are separated from the vortex-free ground state by $\Delta E^s$ and $2\Delta E^s$, respectively, when the two pairs are well separated. At large $s$ the first excited state above both of these configurations is given by the fermion gap to excite a free $\psi$. Since this energy is constant on all vortex-configurations, these levels are located at $\Delta E^s + \Delta_0$ and $2\Delta E^s + \Delta_0$. As $\Delta_0 > \Delta E^s$ the low energy behavior consists only of vortices. Since the energy gap $\Delta_0$ to excite a $\psi$ does not depend on the underlying vortex configuration, this observation generalizes to any configuration where the vortices are kept well separated.

This simple spectral behavior is lost when the vortices are near each other. There the energy levels are shifted and the degeneracies are partially lifted. The smallest separation we can consider is $s=0$, which corresponds to the vortices occupying neighboring plaquettes. However, when the vortices were superposed, they would fuse to the vacuum or to a fermion according to the fusion rules \rf{fusion}, and the spectrum would correspond to the purely fermionic spectrum of the vortex-free sector. This can be connected to the lifting of the degeneracies at small $s$ due to different number of occupied zero modes, i.e. different amount of $\psi$'s in the fusion channels. We observe that at $s=0$ the ground states of both 2- and 4-vortex sectors (no occupied zero modes) tend towards the vortex-free ground state, whereas all the states with occupied zero modes tend to higher energies which correspond to fermionic excitations. This is in agreement with the predictions of the Ising anyon model.

\section{Numerical experiments on finite toroidal lattices}
\label{Numerics}

In this section we present results of a numerical study of different finite size configurations.  While the physics of these compactified systems is often complicated due to finite size effects, they can be used to directly compare numerical and analytical data. After a brief introduction into the methodology used in the numerical calculations, we present a comparison of numerical and analytical data and we show that they are in exact agreement, validating the presented theory. We continue with a more general numerical examination of finite toroidal systems. These studies go beyond the currently available analytical results and illustrate the non-trivial dependence of the low-energy spectrum on the $K$-term.

\subsection{Systems of interest}
\label{systems}

\begin{figure}[t] 
       \centering
       \includegraphics[width=.9\textwidth,height=0.17\textwidth]{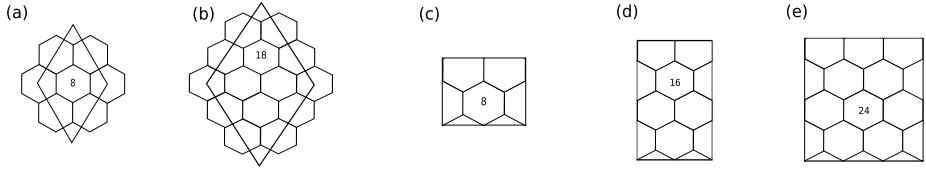}
       \caption{Toroidal lattices used in numerical study. The diamond lattices (a) and (b), are symmetric with respect to exchange of $x,y$ and $z$ links. The rectangular lattices, (c), (d) and (e) are symmetric with respect to $x$ and $y$-links only.} 
       \label{fig:Allconfig}
\end{figure}

Our numerical experiments focus on calculating the low-energy spectral properties of finite-size honeycomb lattice systems with the Hamiltonian given by (\ref{H}). We use a variety of toroidal systems which differ by the total number of spins and by lattice compactification. The size of the system varies from $N = 8$ spins, which constitute an elementary unit of the model that can be compactified on a torus, to N = 24 spins. Though small scale $(N < 18)$ computations require modest computing resources and are conveniently carried out using high level languages (e.g. Matlab), the dimensionality of the lattice Hilbert space scales exponentially with the number of spins (e.g. for $24$ spins, the dimensionality of the Hilbert space is $2^{24} \approx 1.6 \times 10^7$) and thus requires optimized parallel processing which will be discussed further below. In most cases the complexity of the computation can be reduced by taking into account the intrinsic symmetries of the model. These symmetries will be discussed in detail elsewhere \cite{Kells}. We use two physically inequivalent types of finite lattice compactifications on a torus which we call `diamond', (see Figures \ref{fig:Allconfig} (a) and (b)) and `rectangular', (see Figures \ref{fig:Allconfig} (c), (d) and (e)). It can be easily seen that some nontrivial closed loops constructed within one compactification type correspond to open strings in the other type of the same size (for example the cases (a) and (c) in Figure \ref{fig:Allconfig}).

\subsection{Methodology}
\label{methodology}

Our numerical methodology consists of three main components: generation of the Hamiltonian \rf{H} and the plaquette operators $\hat{w}_p$ \rf{w}, their diagonalization and the analysis of the obtained eigenstates $\ket{n}$ and the eigenvalues $E_n$ .

For spin systems with $(N \ge 18)$ we distribute the matrices over a number of different processors using the PETSc library \cite{PETSc1,PETSc2,PETSc3}. The matrix loading routine requires the user to calculate the position and value of the non-zero elements of a particular row of the matrix under consideration. Using this method we can easily store matrices of dimension $2^{24}$. For basic linear algebra routines we use the Linear Algebra Wrapper (LAW) library \cite{LAW}. For the numerical diagonalization of these matrices that are distributed across multiple processors we use the SLEPc library \cite{SLEPc} which is built upon PETSc. The software contains a number of exact diagonalization routines including the ARPACK Arnoldi library \cite{ARPACK,Arnoldi} and an optimized implementation of Krylov-Shur algorithm \cite{Stewart1,Stewart2}. Using Krylov-Shur, and with the matrix distributed across 64 processing nodes, SLEPc returns the lowest 10 energy eigensolutions of the full 24-spin system in under one hour.   
   
The aim of the numerical analysis is to classify the energy eigenvectors $\ket{n}$ according to their vortex configuration. Since all plaquettes commute with the Hamiltonian all energy eigenvectors $\ket{n}$ must satisfy $w_p =\bra{n} \hat{w_p} \ket{n} =\pm1$ for all $N/2$ plaquettes on a torus. However, the relation $\prod_p \hat{w}=I$ in (\ref{w}) implies that there are only $N/2 -1$ independent quantum numbers, $\{ w_1, ...., w_{N/2 -1} \}$ , for each vortex configuration sector. Therefore, in order to reduce the Hilbert space to particular vortex configuration, one must only impose $N/2 -1$ constraints.  Since the imposition of each constraint reduces the dimension of the Hilbert space by a factor of 2 we see that there are only $2^{N/2 -1}$ unique vortex configuration sectors, each with a Hilbert space dimension of $2^{N/2 +1}$.  

In what follows we are only concerned with classifying the sectors according to the total number of vortices. For that we define the vortex counting operator
\begin{equation}
\hat{v} \equiv \frac{1}{4} \left(N \hat{I} -2 \sum_p \hat{w}_p \right).
\end{equation}
The number of vortices corresponding to the eigenstate $\ket{n}$ is given by the expectation value $\bra{n}\hat{v} \ket{n}$.

\subsection{Comparison of numerical and analytical data}
\label{comparison}

In this section we compare results obtained from exact numerical diagonalization with the solutions obtained by the method of Majorana fermionization outlined in Section \ref{periodic}. For the purpose of this comparison we analyze the $N=8$ spin diamond system (see Figure \ref{fig:Allconfig}(a)), which corresponds to the $(2,2)$-unit cell when we employ the Majorana fermionization. Placing this unit cell on a torus implies that all the couplings in the Hamiltonian \rf{H_majorana} are now between sites belonging to the unit cell. In terms of the elements \rf{A12} - \rf{A22} of the matrix $A_{\lambda k \mu l}$ this means setting $\mathbf{v}_x = \mathbf{v}_y = 0$ everywhere.

The $(2,2)$-unit cell contains 12 links on which one must specify the values of the $u$'s. This implies that there are altogether $2^{12} = 4096$ distinct ways to create vortex configurations. However, due to finite size effects there is no a priori way to tell which configuration of the $u$'s will correspond to the lowest ground state energy $E_0$. We carry out a systematic investigation by diagonalizing the resulting $8 \times 8$ Hamiltonian for all the $u$ configurations and find that there are in total 512, 3072 and 512 ways to create 0-vortex, 2-vortex and 4-vortex configurations, respectively. The lowest energy is found to correspond to the 4-vortex configuration and the first excited states to vortex-free configurations.

\begin{figure}[t]
\centerline{
\begin{tabular}{cc}
\epsfxsize=7cm \epsfysize=5cm \epsfbox{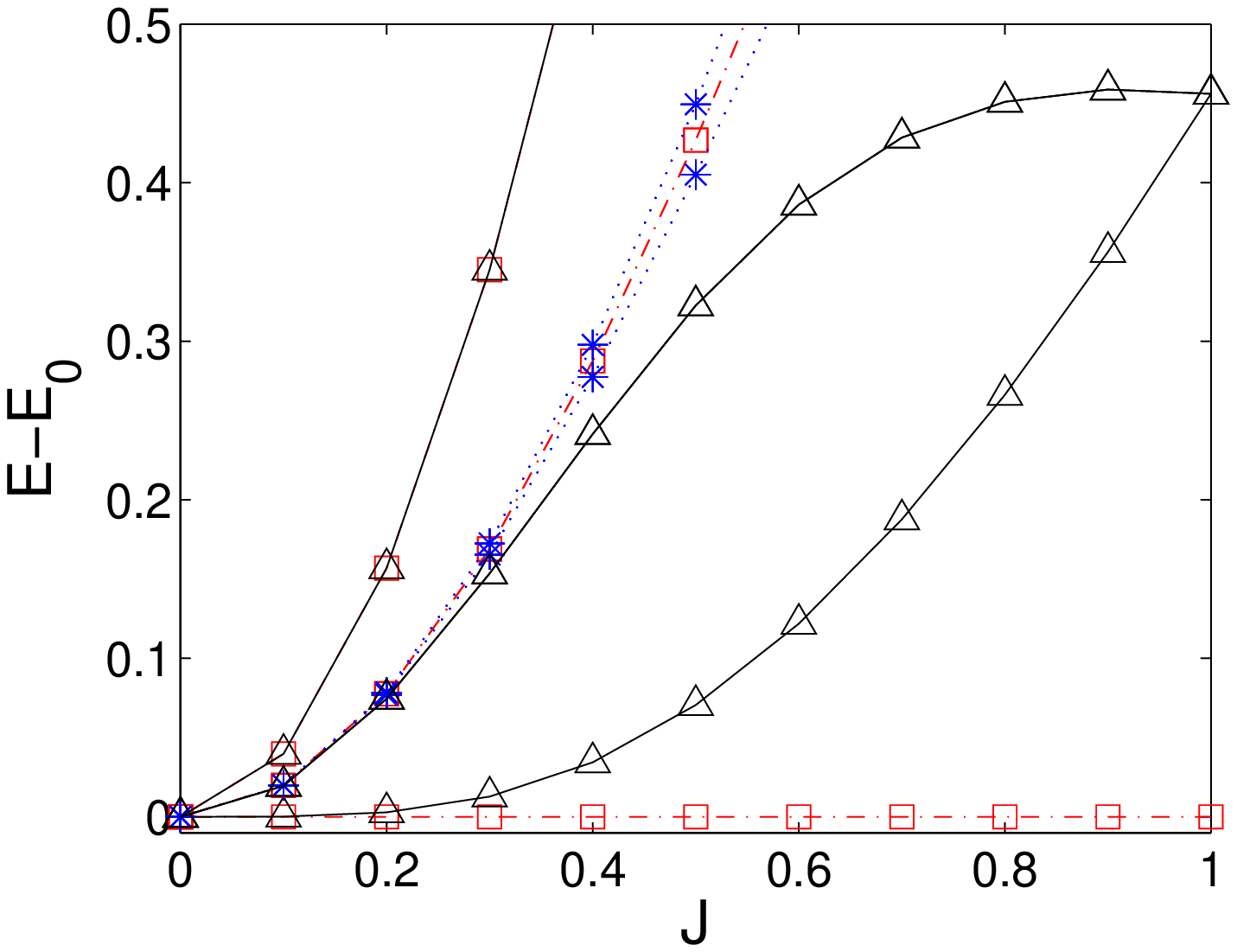} &
\epsfxsize=7cm \epsfysize=5cm \epsfbox{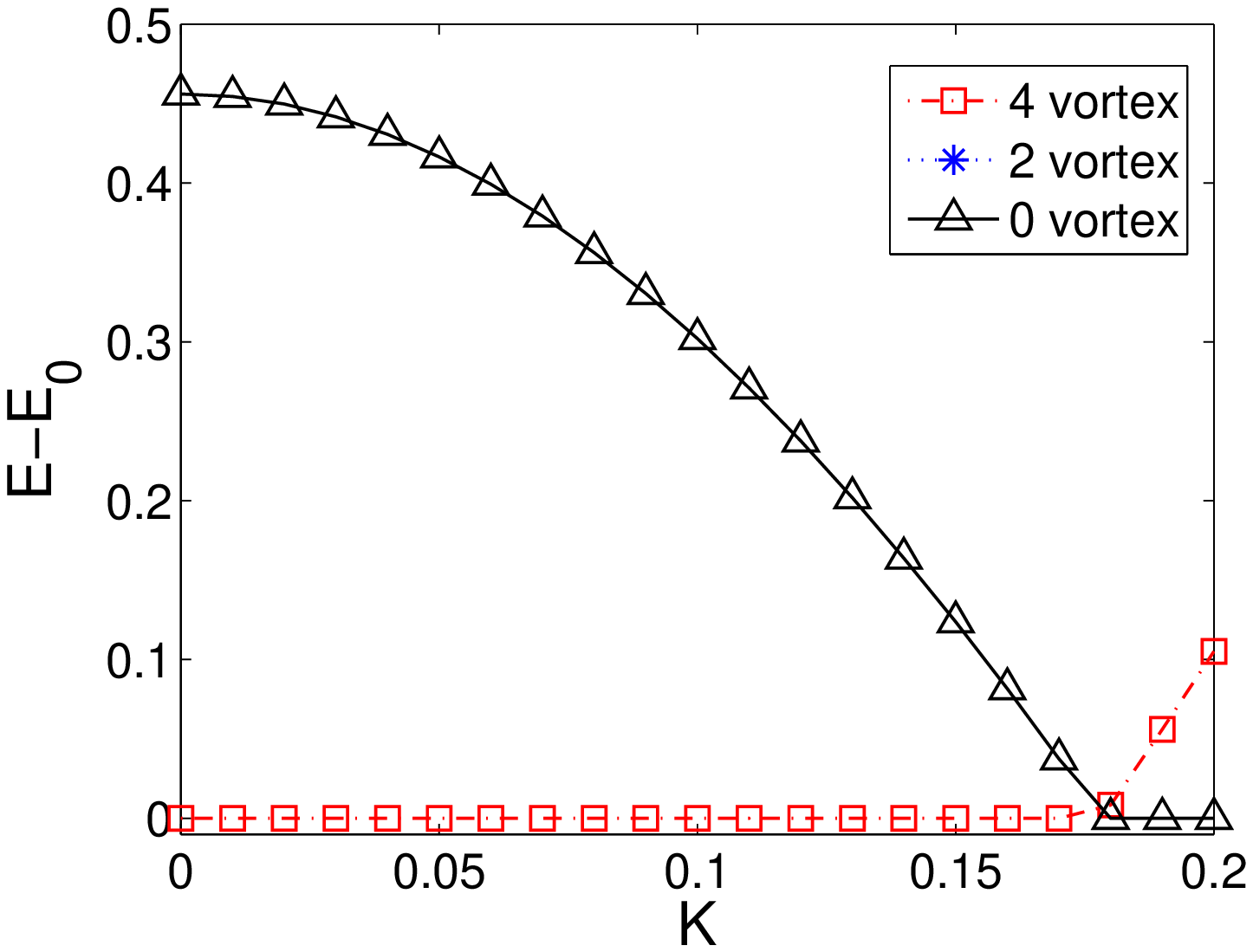} \\ (a) $K=0$ & (b) $J=1$ and $K \geq 0$ \end{tabular} }
\caption{\label{fig:D8} \small{Some of the lowest lying energy eigenvalues for the 8-spin diamond system (see Figure \ref{fig:Allconfig}(a)). (a) $K = 0$, (b) $J=1$ and $K \geq 0$. }}
\end{figure}

Using direct numerical diagonalization we find a ground state, which corresponds also to a 4-vortex configuration. In Figure \ref{fig:D8} we plot some of the lowest lying excited states for (a) $K=0$ and (b) $J=1$ and $K \ge 0$. The values agree with the analytical data to 14 decimal places. In addition, we observe using both numerical and analytic approach the level crossing due to the $K$-term between 0-vortex and 4-vortex sectors (see Figure \ref{fig:D8}(b)), as well as that at $J=1$ the vortex-free ground state is threefold degenerate.

\subsection{Numerical study of the $K$-term}
\label{3body}

In this section we numerically calculate the effect of the K-term on the energy spectrum of the non-abelian B phase for three different finite size lattices. Analysis of the spectrum in all cases shows that the K-term is capable of inducing level crossings between eigenvectors from the same vortex configuration sector. This observation means that the K-term is, in the language of Section \ref{periodic}, capable of opening and closing of fermionic gaps.
 
\begin{figure}[t]
\centerline{
\begin{tabular}{ccc}
\epsfxsize=4.5cm \epsfysize=3.8cm \epsfbox{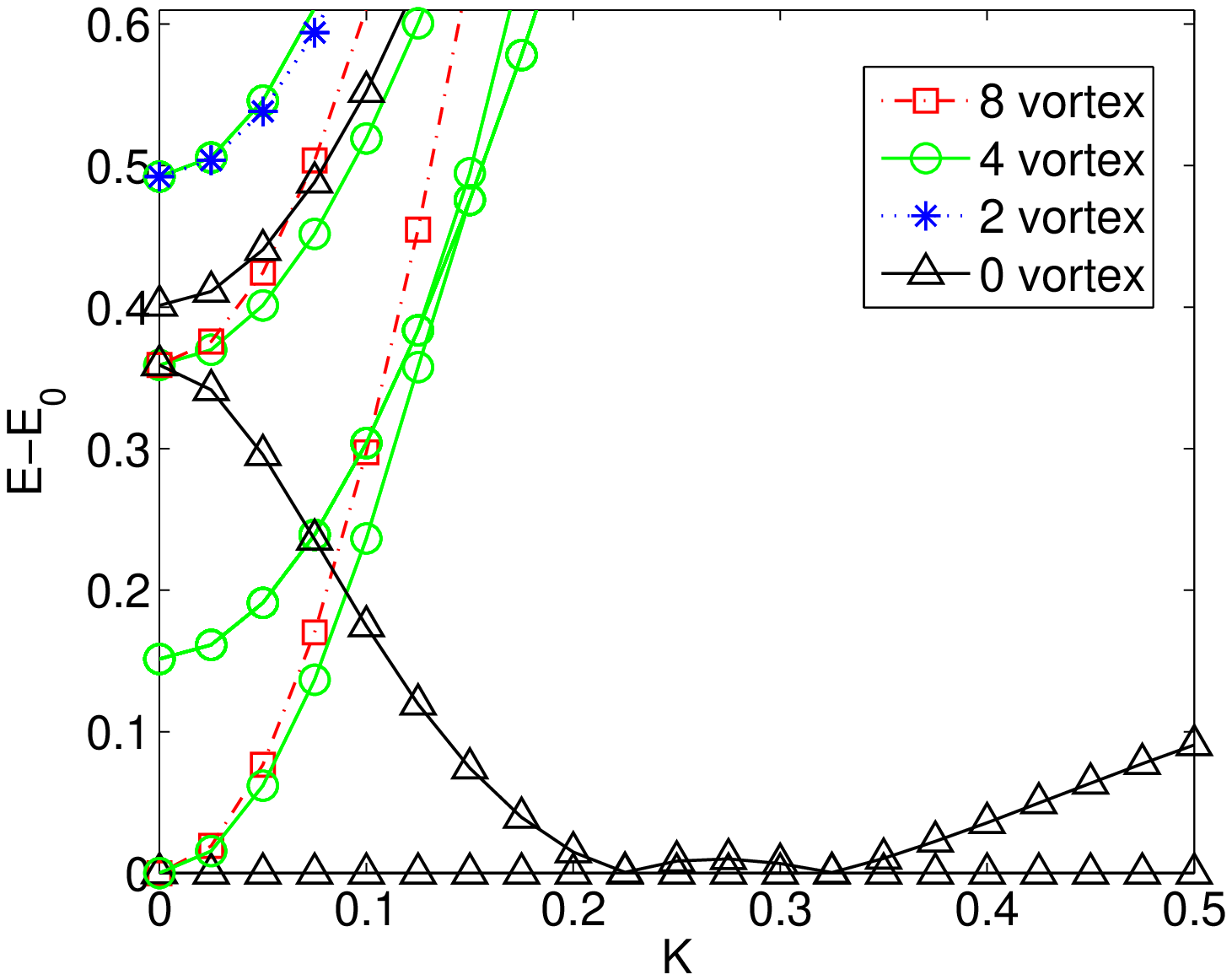} &
\epsfxsize=4.5cm \epsfysize=3.8cm \epsfbox{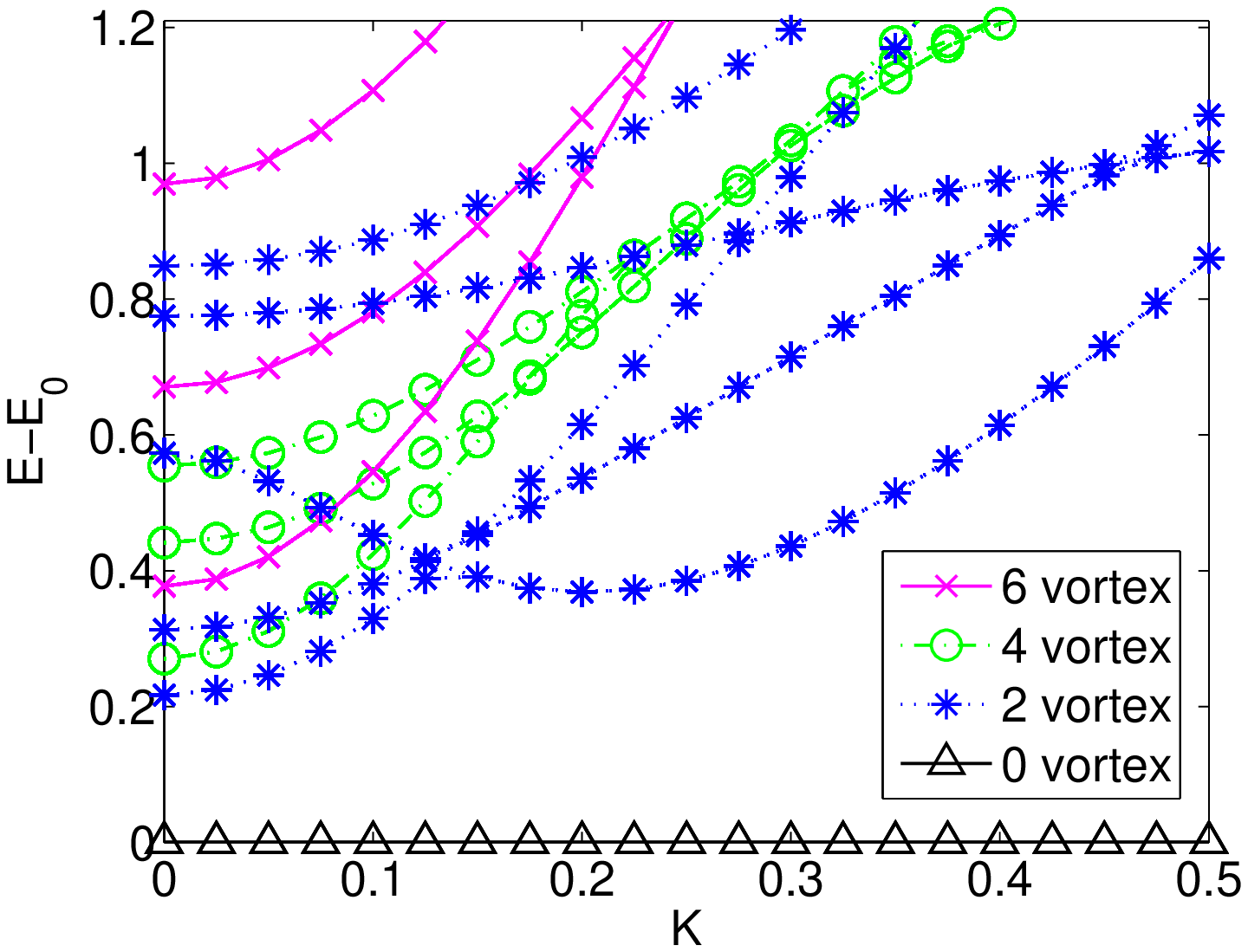} & 
\epsfxsize=4.5cm \epsfysize=3.8cm \epsfbox{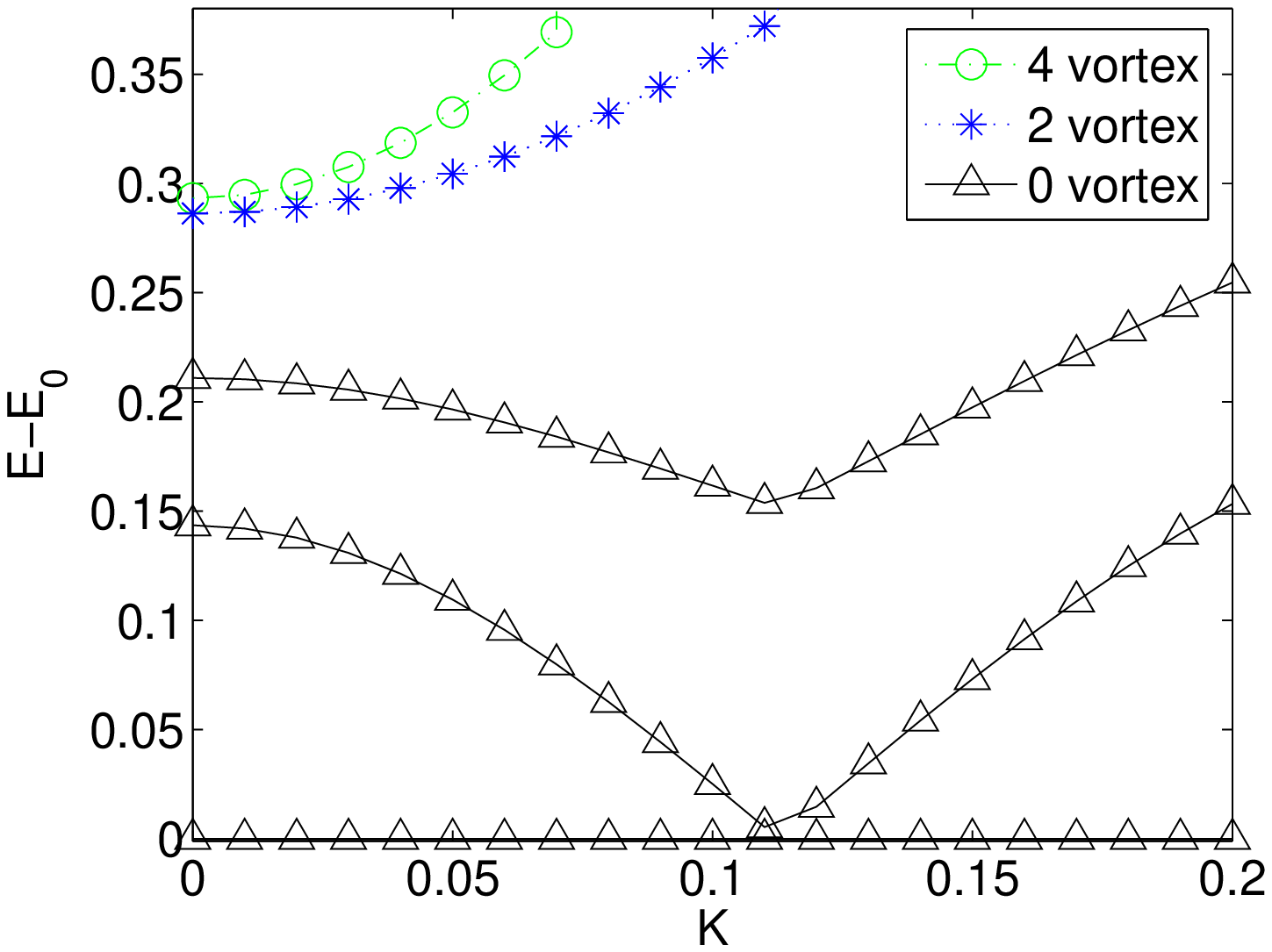} \\ (a) 16-spin & (b) 18-spin & (c) 24-spin \end{tabular} }
\caption{\label{fig:Eall} \small{Some of the lowest lying energy eigenvalues $E$ above the ground state $E_0$ for 16, 18 and 24-spin lattices at $J=1$ and $K \geq 0$. (a) Spectrum of the 16 spin rectangular lattice of Figure \ref{fig:Allconfig}(d). (b) Spectrum of the 18 spin diamond lattice of Figure \ref{fig:Allconfig}(b). (c) Spectrum of the 24 spin rectangular lattice shown in Figure \ref{fig:Allconfig}(e).}}
\end{figure}

We first consider the 16-spin rectangular lattice shown in Figure \ref{fig:Allconfig}(d). The system is symmetric only with respect to reflection in the vertical ($z$-link) axis.  At $J=1$, $K = 0$  the ground state is four times degenerate, containing single 0- and 8-vortex states and two 4-vortex states. The two 4-vortex states are related by a lattice translation.  In Figure \ref{fig:Eall} (a) we plot how the addition of the K-term affects the low energy spectrum. We observe a lifting of the ground state degeneracy, but the 4-vortex states remain still degenerate. We observe also that the K-term can induce spectral crossings between states from the same vortex configuration as seen in the double crossing of the 0-vortex ground states between $K\approx 0.22$ and  $K\approx 0.32$.

In Figure \ref{fig:Eall}(b) we consider the spectrum of 18-spin diamond lattice shown in Figure \ref{fig:Allconfig}(b). It is symmetric with respect to exchange of all $x,y$ or $z$ links and it contains an odd number of plaquettes. At the exact centre of the $B$ phase ($J = 1$) the ground state contains three degenerate states belonging to the 0-vortex sector. We observe that the K-term does not lift the ground state degeneracy. This is to be expected as the $K$-term is also symmetric with respect to exchange of $x,y$ or $z$ links.  However, in general the K-term {\em does} affect the relative energy levels of non-degenerate states of the same vortex sector. This can be seen as the level crossings of the excited states.

Finally, we investigate the 24-spin rectangular lattice shown in Figure \ref{fig:Allconfig}(e). We see that the lowest three states are non-degenerate and belong to the 0-vortex sector. Again we observe the non-trivial behaviour of the spectrum as the function of $K$. Figure \ref{fig:Eall}(c) shows that the as $K$ increases the gap between ground state and first excited state closes and a level crossing occurs at $K \approx 0.11$.

\section{Conclusions}
\label{conclusions}

We have presented an extensive analysis of the spectrum in Kitaev's honeycomb lattice model
\cite{Kitaev05} with the focus on the properties of the non-abelian regime.
Due to the exact solvability of the model we were able to analytically determine,
qualitatively as well as quantitatively, the spectral behavior in the presence of vortices in the thermodynamic limit. This behavior was subsequently identified with the qualitative predictions derived in the context of a $p$-wave superconductor, a model to which the honeycomb lattice model is known to be equivalent \cite{Chen,Yu}. The validity of our results is supported by exact
numerical diagonalizations of various finite size lattice Hamiltonians. There is an exact agreement with the results obtained through the analytical methods. This is a strong validation of the employed
analytical techniques and of the conclusions drawn from them. 

To be precise, our study allowed us to directly compare the spectral behavior in the abelian and in the non-abelian phases and extract characteristics that are unique to the non-abelian phase. The crucial difference is the ground state degeneracy in the non-abelian phase in the presence of well separated vortices. We explain this in terms of zero modes in the spectrum and provide a direct verification that the number of zero modes present in a system with $2n$ vortices is $n$ \cite{Ivanov,Stern,Stone}. The resulting $2^n$-fold degeneracy of the ground state is in agreement with
the non-abelian character of the Ising non-abelian anyons. Furthermore, we observe directly the lifting of the ground state degeneracy when the vortices are brought close to each other and explain the lifting in terms of the fusion rules of the Ising vortices. The energy splitting at short ranges could in principle be used to distinguish the different possible fusion channels without the need to employ an interference procedure. Also, the fact that the information about the fusion outcome is a non-local property of the vortex pair is explicitly demonstrated as the degeneracy present at large $s$. 

Moreover, we have demonstrated that the phase boundary between the abelian and non-abelian phase depends on the underlying vortex configuration and that vortices are interacting in both phases. This attractive interaction is strong in the non-abelian and weak in the abelian phase. Also, the energy gap to excite a pair of vortices is considerably larger in the non-abelian phase. Another characteristic of the non-abelian phase is that the fermionic gap to excite free fermions, $\Delta_0$, does not depend on the underlying vortex configuration as long as the vortices are kept well separated. This means that all the sectors of the non-abelian phase are equally stable with respect to fermionic excitations. Based on this we defined the vortex gap, $\Delta E_{2v}$, in an asymptotic fashion. This gap provides a stability criterion for a particular sector. The combination of all these observations allowed us to outline the low energy spectrum of the non-abelian phase for configurations where the vortices are both well separated and close to each other. We observe that at large separations the spectral behavior consists only of vortices, which is in agreement with the prediction that the low energy behavior should be fully captured by the Ising anyon model. Understanding in detail the behavior of the energy spectrum is of importance to the proposed implementations of this model in the laboratory \cite{Micheli05}. Our observation of the phase boundary dependence on the vortex density for particular values of  $J_{\alpha}$ and $K$ could be of interest to these proposals. In particular, if an increase in temperature is accompanied by an increase in the number of vortices, then it would induce a transition from the non-abelian to the abelian phase.

Finally, we have presented a numerical study of the effect of the $K$-term on the spectra of various finite size systems. These systems are too small to observe behavior similar to the thermodynamic limit, but one important qualitative similarity exists. We observe that the $K$-term is capable
of inducing level crossings of states belonging to the same vortex sector.
This is the finite size equivalent to the opening and closing of fermionic gaps. A more detailed study on these finite size effects will be presented elsewhere \cite{Kells}.

\section*{Acknowledgements}

We would like to thank Joost Slingerland for inspiring conversations and
Kavli Institute of Theoretical Physics and Aspen Centre for Physics for their hospitality where part of this
work was carried out. This work was partially supported by SCALA, EMALI, EPSRC, the Royal
Society, the Finnish Academy of Science, the Science Foundation Ireland and the Irish Centre for High-End Computing.

\end{document}